\documentclass[%
 reprint,
 superscriptaddress,
 preprintnumbers,
 nofootinbib,
 amsmath,amssymb,
 aps,
 prd,
 floatfix,
]{revtex4-2}

\usepackage{aas_macros}

\usepackage{graphicx}
\usepackage{xcolor}
\definecolor{plasmablue}{rgb}{0.050383, 0.029803, 0.527975}

\makeatletter
\def\Dated@name{}
\makeatother

\usepackage{hyperref}
\hypersetup{
  colorlinks=true,
  linkcolor=plasmablue,
  urlcolor=plasmablue,
  citecolor=plasmablue,
  pdftitle={SPT Clusters with DES and HST Weak Lensing. II. Cosmological Constraints from the Abundance of Massive Halos},
  pdfauthor={Bocquet et al.},
  }

\defcitealias{bocquet24I}{Paper~I}

\renewcommand{\vec}[1]{\mbox{\boldmath$#1$}}
\newcommand{\asz}{\ensuremath{{A_\mathrm{SZ}}}}
\newcommand{\bsz}{\ensuremath{{B_\mathrm{SZ}}}}
\newcommand{\csz}{\ensuremath{{C_\mathrm{SZ}}}}
\newcommand{\sigmalnzeta}{\ensuremath{{\sigma_{\ln\zeta}}}}
\newcommand{\alambda}{\ensuremath{{A_\lambda}}}
\newcommand{\blambda}{\ensuremath{{B_\lambda}}}
\newcommand{\clambda}{\ensuremath{{C_\lambda}}}
\newcommand{\sigmalnlambda}{\ensuremath{{\sigma_{\ln\tilde\lambda}}}}

\newcommand{\Mwl}{\ensuremath{M_\mathrm{WL}}}
\newcommand{\Mhalo}{\ensuremath{M_\mathrm{halo}}}

\newcommand{\dif}{\mathrm{d}}

\newcommand{\Obhh}{\ensuremath{\Omega_\mathrm{b}h^2}}
\newcommand{\Om}{\ensuremath{\Omega_\mathrm{m}}}

\newcommand{\Onuhh}{\ensuremath{\Omega_\nu h^2}}
\newcommand{\sig}{\ensuremath{\sigma_8}}
\newcommand{\Sopt}{\ensuremath{S_8^\mathrm{opt}}}
\newcommand{\LCDM}{\ensuremath{\Lambda\mathrm{CDM}}}
\newcommand{\wCDM}{\ensuremath{w\mathrm{CDM}}}
\newcommand{\sumMnu}{\ensuremath{\sum m_\nu}}

\newcommand{\Msun}{\ensuremath{\mathrm{M}_\odot}}

\begin{document}

\preprint{DES-2023-787}
\preprint{FERMILAB-PUB-23-522-PPD}

\title{SPT Clusters with DES and HST Weak Lensing. II. Cosmological Constraints from the Abundance of Massive Halos}

\author{S.~Bocquet}
\email{sebastian.bocquet@physik.lmu.de}
\affiliation{University Observatory, Faculty of Physics, Ludwig-Maximilians-Universit\"at, Scheinerstr. 1, 81679 Munich, Germany}
\author{S.~Grandis}
\affiliation{Universit\"at Innsbruck, Institut f\"ur Astro- und Teilchenphysik, Technikerstr. 25/8, 6020 Innsbruck, Austria}
\affiliation{University Observatory, Faculty of Physics, Ludwig-Maximilians-Universit\"at, Scheinerstr. 1, 81679 Munich, Germany}
\author{L.~E.~Bleem}
\affiliation{High-Energy Physics Division, Argonne National Laboratory, 9700 South Cass Avenue, Lemont, IL 60439, USA}
\affiliation{Kavli Institute for Cosmological Physics, University of Chicago, 5640 South Ellis Avenue, Chicago, IL 60637, USA}
\author{M.~Klein}
\affiliation{University Observatory, Faculty of Physics, Ludwig-Maximilians-Universit\"at, Scheinerstr. 1, 81679 Munich, Germany}
\author{J.~J.~Mohr}
\affiliation{University Observatory, Faculty of Physics, Ludwig-Maximilians-Universit\"at, Scheinerstr. 1, 81679 Munich, Germany}
\affiliation{Max Planck Institute for Extraterrestrial Physics, Gie{\ss}enbachstr.~1, 85748 Garching, Germany}
\author{T.~Schrabback}
\affiliation{Universit\"at Innsbruck, Institut f\"ur Astro- und Teilchenphysik, Technikerstr. 25/8, 6020 Innsbruck, Austria}
\affiliation{Argelander-Institut f\"ur Astronomie, Auf dem H\"ugel 71, 53121 Bonn, Germany}

\author{T.~M.~C.~Abbott}
\affiliation{Cerro Tololo Inter-American Observatory, NSF's National Optical-Infrared Astronomy Research Laboratory, Casilla 603, La Serena, Chile}
\author{P.~A.~R.~Ade}
\affiliation{School of Physics and Astronomy, Cardiff University, Cardiff CF24 3YB, United Kingdom}
\author{M.~Aguena}
\affiliation{Laborat\'orio Interinstitucional de e-Astronomia - LIneA, Rua Gal. Jos\'e Cristino 77, Rio de Janeiro, RJ - 20921-400, Brazil}
\author{A.~Alarcon}
\affiliation{High-Energy Physics Division, Argonne National Laboratory, 9700 South Cass Avenue, Lemont, IL 60439, USA}
% ...

\collaboration{the SPT and DES Collaborations}
\noaffiliation

\date{Phys. Rev. D accepted 24 May 2024}

\begin{abstract}
We present cosmological constraints from the abundance of galaxy clusters selected via the thermal Sunyaev-Zel'dovich (SZ) effect in South Pole Telescope (SPT) data with a simultaneous mass calibration using weak gravitational lensing data from the Dark Energy Survey (DES) and the \textit{Hubble Space Telescope} (HST). The cluster sample is constructed from the combined SPT-SZ, SPTpol~ECS, and SPTpol~500d surveys, and comprises 1,005 confirmed clusters in the redshift range $0.25-1.78$ over a total sky area of 5,200~deg$^2$. We use DES~Year~3 weak-lensing data for 688~clusters with redshifts $z<0.95$ and HST weak-lensing data for 39 clusters with $0.6<z<1.7$. The weak-lensing measurements enable robust mass measurements of sample clusters and allow us to empirically constrain the SZ observable--mass relation without having to make strong assumptions about, e.g., the hydrodynamical state of the clusters. For a flat $\Lambda$CDM cosmology, and marginalizing over the sum of massive neutrinos, we measure $\Omega_\mathrm{m}=0.286\pm0.032$, $\sigma_8=0.817\pm0.026$, and the parameter combination $\sigma_8\,(\Omega_\mathrm{m}/0.3)^{0.25}=0.805\pm0.016$. Our measurement of $S_8\equiv\sigma_8\,\sqrt{\Omega_\mathrm{m}/0.3}=0.795\pm0.029$ and the constraint from {\it Planck} CMB anisotropies (2018 TT,TE,EE+lowE) differ by $1.1\sigma$. In combination with that {\it Planck} dataset, we place a 95\% upper limit on the sum of neutrino masses $\sum m_\nu<0.18$~eV. When additionally allowing the dark energy equation of state parameter $w$ to vary, we obtain $w=-1.45\pm0.31$ from our cluster-based analysis. In combination with {\it Planck} data, we measure $w=-1.34^{+0.22}_{-0.15}$, or a $2.2\sigma$ difference with a cosmological constant. We use the cluster abundance to measure $\sigma_8$ in five redshift bins between 0.25 and 1.8, and we find the results to be consistent with structure growth as predicted by the $\Lambda$CDM model fit to {\it Planck} primary CMB data.
\end{abstract}

\maketitle

\section{Introduction}

Understanding the cause for the accelerated expansion of our universe is currently one of the biggest challenges in physics.
To address this robustly, a multitude of different probes and techniques have been proposed and are being used to observationally constrain the cosmological parameters.
Interestingly, the comparison of results obtained from these probes have recently uncovered two anomalies: First, the present-day value of the Hubble parameter determined from primary anisotropies of the cosmic microwave background (CMB) at high redshift is in at least $4\sigma$ tension with measurements in the local universe using the distance ladder e.g., \cite{riess18, planck18VI, riess20, riess22}.
Second, measurements of the $S_8$-parameter\footnote{$S_8\equiv\sigma_8\sqrt{\Om/0.3}$ is a combination of the amplitude of fluctuations in the linear matter density field on scales of $8\,h^{-1}$Mpc ($\sigma_8$) and the matter density \Om.} by low-redshift probes have tended to be lower than predicted by the $\Lambda$CDM model with parameters set by high-redshift CMB data e.g., \cite{DESY13x2pt, vanuitert18, bocquet19, heymans21, DESY33x2pt, kobayashi22, dalal23hsc, li23hsc, DESY3KiDS1000}.

Over recent years, new large-scale structure probes such as joint analyses of galaxy clustering and weak-lensing 2-point correlation functions (so-called 3$\times$2~pt analyses) \citep{DESY33x2pt, heymans21, more23}, the CMB lensing power spectrum \citep{planck18VIII, bianchini20, qu24}, or the combination of all of these \cite{chang23} have emerged.
Measurements of the abundance of massive dark matter halos (and the galaxy clusters they host) probe the matter density field through a different mechanism and are subject to different systematics, thus offering an ideal and powerful complement \citep{haiman01, dodelson16}.
However, for a study of halo abundance to be successful, the relationship between the underlying, un-observable halo mass and the observable properties of the clusters needs to be known (see e.g., reviews \cite{allen11, pratt19}).

In this work, we present cosmological constraints derived from a sample of 1,005~galaxy clusters detected using the South Pole Telescope (SPT) \citep{carlstrom11} in combination with galaxy and weak-lensing data from the Dark Energy Survey (DES) \citep{flaugher15, DES16, DES18DR1}, photometric data from the Wide-field Infrared Survey Explorer (WISE) \cite{WISEobservatory}, and targeted weak-lensing measurements from the {\it Hubble Space Telescope} (HST).
We select the galaxy clusters via the thermal Sunyaev-Zel'dovich (hereafter SZ) \citep{sunyaev&zeldovich72} effect in data from the first two SPT cameras, SPT-SZ and SPTpol (SPT-3G being the third, currently operating camera).
The SZ effect is sourced by the electron pressure in the hot gas component of the cluster (the intra-cluster medium, or ICM) and arises when CMB photons scatter with these high-energy electrons, leading to a distinct spectral signature.
The SZ effect signal is thus a clean tracer of the hot ICM and enables the detection of galaxy clusters out to the highest redshifts at which massive halos exist, thereby creating an essentially mass-limited sample whose limiting mass does not evolve strongly with redshift.
Optical and near-infrared data from DES, WISE, and other targeted observation programs are then used to confirm clusters (and to exclude false detections) and to assign redshifts to all confirmed systems.
With these follow-up data taken from wide-field surveys, we can robustly calibrate the probability of chance association of random optical structures with an SZ noise fluctuations through measurements along random lines of sight \citep{klein18, bleem20}.

To relate the strength of the SZ effect signal to the underlying halo mass, we cannot rely on first principles because modeling the dynamical state of the ICM is highly complex.
In particular, the ICM cannot be assumed to be in hydrostatic equilibrium.
Instead, we use measurements of the weak gravitational shear, which depends on the total halo mass irrespective of its dynamical state, and which can thus be mapped to halo mass with excellent control over systematic uncertainties (for a review on cluster lensing, see \cite{umetsu20}).
We now leverage the strategic overlap of the SPT survey and DES footprints and---for the first time---use DES lensing data for mass calibration in a cosmological analysis of SPT-selected clusters.
Within our modeling framework, we empirically constrain the SZ--mass relation in a robust, weak-lensing-based way, thereby addressing and overcoming the main challenge in cluster cosmology: the accurate calibration of cluster masses.

Using 5,200~deg$^2$ of SPT cluster surveys, we select 1,005~clusters at redshift $z>0.25$.
In the 3,567~deg$^2$ overlap region between the DES and SPT surveys, we obtain measurements of weak-lensing shear from DES Year~3 data (hereafter DES~Y3) for 688~clusters.
Since DES lensing runs out of constraining power for high-redshift ($z\gtrsim0.9$) clusters, we supplement the DES lensing dataset with targeted lensing observations of 39 clusters in the redshift range $0.6<z<1.7$ using the HST.
This joint cluster and lensing dataset represents a significant improvement over the previous analysis of SPT clusters \cite{bocquet19}, which was based on 344~clusters in the SPT-SZ survey \cite{bleem15}.
In that analysis, we used cluster lensing data from targeted programs at Magellan (19 clusters, \cite{dietrich19}) and HST (13 clusters, \cite{schrabback18}).
The reduced statistical and systematic uncertainties of the dataset we consider here enables significantly improved cosmological constraints.

In this paper, we present cosmological constraints from the joint analysis of the SPT cluster abundance with a weak-lensing mass calibration using DES and HST data. The data products, the cluster lensing measurements using DES~Y3 data, the modeling choices, the analysis pipeline, and its validation are presented in detail in a companion paper (\cite{bocquet24I}, hereafter \citetalias{bocquet24I}).
We briefly review the dataset in Sec.~\ref{sec:data}.
We summarize the analysis method in Sec.~\ref{sec:method}.
We discuss our blinding strategy and the robustness tests that were performed prior to unblinding of the results in Sec.~\ref{sec:robustness_tests}.
We present our cosmological constraints in Sec.~\ref{sec:results}, and close with a summary in Sec.~\ref{sec:summary}.

Throughout this analysis, we assume spatial flatness.
The parameter combination $S_8\equiv\sigma_8\left(\Om/0.3\right)^{0.5}$ is defined to be tightly constrained by cosmic shear.
We empirically determine that our cluster dataset optimally constrains a parameter combination with an exponent of 0.225.
However, for comparability, we adopt $\Sopt\equiv\sigma_8\,\left(\Om/0.3\right)^{0.25}$, which is the optimal combination for CMB lensing \citep{qu24}.\footnote{The uncertainties in our constraints on $\sigma_8\,\left(\Om/0.3\right)^{0.225}$ and $\sigma_8\,\left(\Om/0.3\right)^{0.25}$ differ by less than a percent.}
Halo masses $M_{200\mathrm{c}}$ refer to the mass enclosed within a sphere of radius $r_{200\mathrm{c}}$, within which the mean density is 200 times larger than the critical density $\rho_\mathrm{c}(z)$ at the cluster redshift $z$.
We express the (multivariate) normal distribution with mean $\vec\mu$ and (co)variance $\mathbf K$ as $\mathcal N(\vec\mu, \mathbf K)$.

\section{Data}
\label{sec:data}

We summarize the data products used in this analysis. For a detailed description, we refer the reader to \citetalias{bocquet24I}.

\subsection{SPT Cluster Catalog}

We use the cluster catalogs from the SPT-SZ, SPTpol~ECS, and SPTpol~500d surveys \citep{bleem15, bleem20, klein23spt, bleem24}.
Combined, they cover over 5,200~deg$^2$ of the southern sky. Note that the deep SPTpol~500d survey lies within the footprint of the SPT-SZ survey; in the overlap area, we use only data from SPTpol~500d.
The cluster candidate list is selected on the basis of SPT detection significance $\xi$ and redshift $z$.

Over the part of the SPT survey that is not covered by DES (27\% of the total SPT-SZ and SPTpol cluster survey area), we confirm clusters and assign redshifts using targeted optical observations (using among others, the PISCO imager \cite{stalder14PISCO}) as described in \cite{bleem15, bleem20}.
We select all candidates with
\begin{equation}
  \label{eq:select_noDES}
  \begin{split}
    \xi&>5, \\
    z&>0.25,
  \end{split}
\end{equation}
which, after follow up, results in a highly pure cluster sample ($\gtrsim95\%$) \citep{bleem15, bleem20}.

\begin{figure*}
  \includegraphics[width=\textwidth]{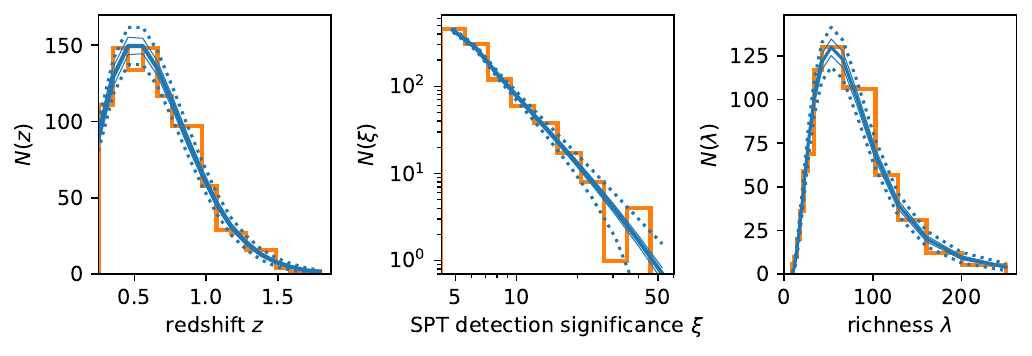}
  \caption{Projection of the 1,005 clusters in the SPT sample along the dimensions of redshift, SPT detection significance, and optical richness.
  Orange histograms show the data.
  Thick lines show the mean recovered model in an analysis of the cluster abundance and cluster lensing, thin solid lines show the $1\sigma$ uncertainty on the mean, and dotted lines show the $1\sigma$ Poisson uncertainty (statistical shot noise).
  There is qualitative agreement between the model and the data.
  }
  \label{fig:dN_dzdxidlambda}
\end{figure*}

Over the 3,567~deg$^2$ of the SPT survey that is covered by DES (73\% of the total survey area), we obtain measurements of optical richness $\lambda$ and of redshift $z$ using the multi-component matched filter cluster confirmation tool (MCMF) \citep{klein18, klein23spt}.
At high cluster redshifts $z>1.1$, which are beyond the reach of DES, we use data from WISE to compute richness and redshift.
To characterize the probability of chance associations, we also compute richness and redshift for random lines of sight in the DES footprint.
With this information, we can define a minimum richness $\lambda_\mathrm{min}$ above which we consider an SZ cluster candidate as confirmed.
In practice, $\lambda_\mathrm{min}$ depends on redshift, and we define $\lambda_\mathrm{min}(z)$ such that the final sample purity is $>98\%$ (see Fig.~2 in \citetalias{bocquet24I}).\footnote{Any remaining level of sample contamination is thus within the shot noise of the sample.}
At fixed SPT detection significance, the different SPT surveys have different levels of purity because of the varying depths.
To keep the overall sample purity approximately constant, over the joint SPT--DES area, we select clusters according to
\begin{equation}
  \label{eq:select_DES}
  \begin{split}
    \xi&>4.25 \,/\, 4.5 \,/\, 5 \,\,(\text{500d / SZ / ECS}), \\
    \lambda&>\lambda_\mathrm{min}(z), \\
    z&>0.25.
  \end{split}
\end{equation}

The final sample comprises 1,005 confirmed clusters with redshift measurements.\footnote{The statistical uncertainties in the redshift measurements do not limit the cosmological constraining power of our cluster dataset.}
All clusters in the DES region also have richness measurements and optically determined center positions (from DES or WISE).
Figure~\ref{fig:dN_dzdxidlambda} shows the distribution of clusters as a function of redshift, SPT detection significance, and optical richness.

\subsection{DES Y3 Weak-Lensing Data}

The DES covers 5,000~deg$^2$ of the southern sky in the $g$, $r$, $i$, $z$, and $Y$ bands.
The DES Y3 cosmology dataset covers 4,143~deg$^2$ after masking, of which 3,567~deg$^2$ overlap with the SPT cluster surveys.
The weak-lensing shape catalog \citep{gatti21} is created by applying the \textsc{Metacalibration} pipeline \citep{huff&mandelbaum17, sheldon&huff17} to data from the $r$, $i$, and $z$ bands.
Detailed information about the photometric dataset \citep{sevilla-noarbe21}, the modeling of the point-spread function \citep{jarvis21}, and image and survey simulations \citep{maccrann22, everett22} can be found in dedicated DES~Y3 publications.
Following the 3$\times$2~pt analysis \citep{DESY33x2pt}, we select lensing source galaxies in four tomographic bins.
The redshift distributions of these source bins are calibrated using self-organizing maps \citep{myles21}.

\begin{figure}
  \includegraphics[width=\columnwidth]{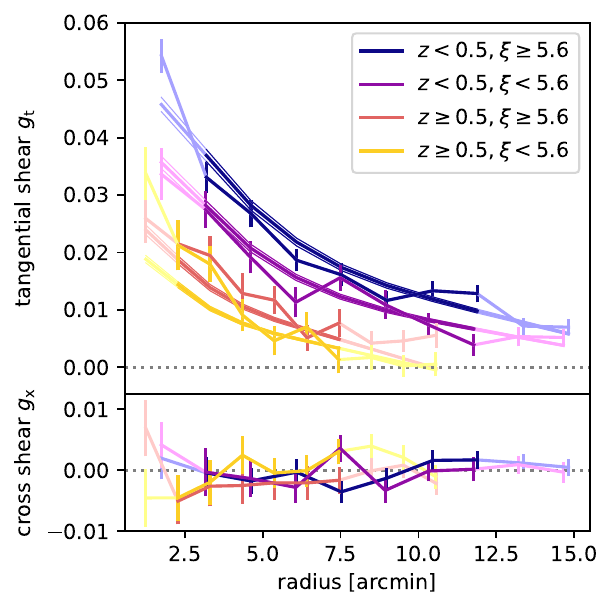}
  \caption{
  Stacked DES~Y3 lensing shear profiles.
  \emph{Upper panel:} Tangential shear, along with the mean model prediction from an analysis of the cluster abundance and cluster lensing.
  The likelihood analysis is performed on a cluster-by-cluster basis, the stacks are only shown for validation purposes.
  The dark shading shows the radial ranges used in the analysis, light shaded data points outside that range are only shown for reference.
  For the 26 data points included in the analysis, we measure a signal-to-noise ratio of 32.2 and we obtain a reduced $\chi^2_\mathrm{red}=1.48$ between the data and the model.
  \emph{Lower panel:} Within the radial ranges used in the analysis, the cross shear is consistent with null with $\chi^2_\mathrm{red}=0.90$.
  }
  \label{fig:stacked_shear}
\end{figure}

For every SPT cluster in the DES footprint, we extract a weak-lensing shear profile within the radial range $0.5<r/(h^{-1}\mathrm{Mpc})<3.2\, (1+z_\mathrm{cluster})^{-1}$, centered on the optically determined cluster center.\footnote{We convert the measured angular separations to physical separations assuming $\Om=0.3$.}
This radial cut avoids the problematic central region of the cluster (affected by feedback from active galactic nuclei, miscentering, blending, cluster member contamination, non-linear shear) and ensures that only the 1-halo term regime is considered \citep{grandis21}. We restrict the use of DES weak-lensing data to clusters below redshift $z=0.95$, which corresponds to the median redshift of the highest-redshift source bin.
We extract 688 cluster shear profiles from a total of 555,912~source galaxies.
For illustrative purposes, we show stacked shear profiles in bins of cluster redshift and SPT detection significance in Fig.~\ref{fig:stacked_shear}.

For the purpose of robustness checks (see Sec.~\ref{sec:robustness_tests}), we also measure shear profiles in a more conservative radial range $0.8<r/(h^{-1}\mathrm{Mpc})<3.2\, (1+z_\mathrm{cluster})^{-1}$.
As another alternative, we measure shear around the cluster centers as determined in the SZ analysis.
We estimate the level of cluster member contamination using two different redshift estimators (\textsc{DNF} \citep{devicente16} as the default, and \textsc{BPZ} \citep{benitez00} as an alternative).

We quantify and discuss the relevant sources of systematic and statistical uncertainties related to the weak-lensing measurements (cluster member contamination, miscentering of the shear profile, shear and photo-$z$ calibration, halo mass modeling, impact of large-scale structure) in detail in \citetalias{bocquet24I}, Sec.~V.

\subsection{High-Redshift HST Weak-Lensing Data}

The DES lensing data does not have substantial constraining power for lenses with $z\gtrsim0.9$, and we use HST data to enable weak-lensing mass calibration at higher redshifts.
Using the HST-39 dataset presented and analyzed in \cite{schrabback18, raihan20, hernandez-martin20, schrabback21, zohren22}, we have a sample of 39 clusters in the redshift range $0.6-1.7$ with space-based weak-lensing measurements.

\section{Analysis Method}
\label{sec:method}

The analysis method is introduced and described in detail in \citetalias{bocquet24I}.
Here, we briefly review the key features.

\subsection{Summary of Analysis Strategy}
\label{sec:strategy}

We set up a Bayesian (hierarchical) population model to describe our dataset (see also \cite{mantz10I, benson13, bocquet15}).
We explicitly model the selection criteria that we adopted to construct the cluster catalog.
Specifically, these are cuts in the SPT detection significance $\xi$, cuts in the optical richness $\lambda$ (over the footprint that is common with DES), and a cut in redshift [see Eqs.~(\ref{eq:select_noDES}) and (\ref{eq:select_DES})].
Clusters with weak-lensing data additionally have measurements of tangential shear profiles.

We follow analytical and simulation-based models for observable--mass relations e.g., \citep{kaiser86, angulo12}, and describe the scaling relation between the unbiased SPT detection significance $\zeta$ and mass as
\begin{equation}
  \begin{split}
    \langle\ln\zeta\rangle =& \ln\asz + \bsz \ln\left(\frac{M_{200\mathrm{c}}}{3\times 10^{14}\,h^{-1}\Msun}\right) \\
    & + \csz \ln\left(\frac{E(z)}{E(0.6)}\right) \label{eq:zetaM}
  \end{split}
\end{equation}
with lognormal intrinsic scatter of width \sigmalnzeta, and relate the detection significance $\xi$ to $\zeta$ as
\begin{equation} \label{eq:xizeta}
P(\xi|\zeta) = \mathcal N\left(\sqrt{\zeta^2 + 3}, 1\right)
\end{equation}
to account for the maximization bias when measuring $\xi$ in noisy data \citep{vanderlinde10}.
We account for the varying depth within the SPT survey by rescaling \asz\ and \csz\ for each individual field \citep{bleem15, bleem20, bleem24}.
For the SPTpol~ECS survey, the overall normalization $\gamma_\mathrm{ECS}$ of \asz\ is difficult to calibrate (owing to detector linearity changes under high atmospheric loading at the elevation of these fields, see Sec.~2.2 \cite{bleem20}), and we thus consider $\gamma_\mathrm{ECS}$ as an additional fit parameter.

We choose the same ansatz for the relation between the intrinsic richness $\tilde\lambda$ and mass
\begin{equation}
  \begin{split}
    \langle\ln\tilde\lambda\rangle =& \ln\alambda + \blambda \ln\left(\frac{M_{200\mathrm{c}}}{3\times10^{14}\,h^{-1}\Msun}\right) \\
    &+ \clambda \ln\left(\frac{1+z}{1.6}\right), \label{eq:lambdaM}
  \end{split}
\end{equation}
with lognormal intrinsic scatter of width \sigmalnlambda, and relate the observed richness $\lambda$ to $\tilde\lambda$ as
\begin{equation}
  \label{eq:P_lambda_tildelambda}
  P(\ln\lambda|\ln\tilde\lambda) = \mathcal N( \ln\tilde\lambda, 1/\tilde\lambda).
\end{equation}
We consider two different types of measured richness; those based on DES data, which we use for clusters with redshift $z<1.1$, and those based on WISE, which we use for clusters with redshifts beyond~1.1.
Rather than attempting to match the two types of richness within an overlap range in redshift, we fit a separate observable--mass relation for each of the two types of richness, with a clean transition from DES-based to WISE-based richness at redshift $z=1.1$.

Using the relations defined in Eqs.~(\ref{eq:zetaM})--(\ref{eq:P_lambda_tildelambda}), we can model the $\xi$--$\lambda$--$z$ sample by convolving the halo mass function with the observable--mass relations (accounting for the covariance in intrinsic scatter $\rho_\mathrm{SZ,\tilde\lambda}\, \sigmalnzeta\, \sigmalnlambda$) and applying the sample selection cuts.
There is no external information on the parameters $A, B, C$ of the scaling relations or on the intrinsic scatter \sigmalnzeta, \sigmalnlambda, and we use weak-lensing data of sample clusters to calibrate these observable--mass relation parameters empirically.
Because we can infer the halo mass from lensing data with excellent control over systematic uncertainties, this is a robust analysis strategy.

We introduce the modeling framework with which we analyze the DES~Y3 lensing data in \citetalias{bocquet24I}.
Here, we briefly review the method.
We adopt a simple model for the projected halo mass distribution $\Sigma$ based on a modified Navarro-Frenk-White profile (NFW) \citep{navarro97}.\footnote{To approximately account for the effect of miscentering, we modify the NFW profile such that it is constant within the typical cluster miscentering radius $R_\mathrm{mis}$ (see \citetalias{bocquet24I}, Sec.~V~B).}
We introduce a latent variable that we call the ``weak-lensing mass'' \Mwl, which we define such that the reduced tangential shear is
\begin{equation}
  \label{eq:g_model}
  g_\mathrm{t}(r,\Mwl) = \frac{\Delta\Sigma(r,\Mwl)~\Sigma_\mathrm{crit}^{-1}}{1-\Sigma(r,\Mwl)~\Sigma_\mathrm{crit}^{-1}}~\bigl(1-f_\mathrm{cl}(r)\bigr).
\end{equation}
The density contrast $\Delta\Sigma(r)\equiv\langle\Sigma(<r)\rangle-\Sigma(r)$ is computed from $\Sigma$, and the lensing efficiency $\Sigma_\mathrm{crit}^{-1}$ is computed from the distribution of source redshifts and the lens redshift.
Note that $\Sigma_\mathrm{crit}^{-1}$ explicitly depends on cosmology.
The cluster member contamination $f_\mathrm{cl}(r)$ accounts for contaminants in the lensing source sample that are not sheared and thus bias the measurement low [therefore, $(1-f_\mathrm{cl})^{-1}$ is often referred to as the boost factor].

The cluster lensing model we just introduced is not perfect (e.g., halos do not individually match the NFW profile) and \Mwl\ inferred from the shear profile $\boldsymbol g_\mathrm{t}$ is thus a biased and noisy estimator of the true halo mass \Mhalo\ \citep{becker&kravtsov11, oguri&hamana11}.
To account for this, we establish a mean relationship,
\begin{equation}
    \left\langle\ln \left(\frac{M_\mathrm{WL}}{M_0}\right)\right\rangle = b_\mathrm{WL}(z) + b_{\mathrm{WL},M} \ln\left(\frac{M_{200\mathrm{c}}}{M_0}\right), \label{eq:MwlM}
\end{equation}
and describe the width of the lognormal scatter around the mean as
\begin{equation}
    \ln\sigma_{\ln M_\mathrm{WL}} = \frac12 \left[s_\mathrm{WL}(z) + s_{\mathrm{WL},M} \ln\left(\frac{M_{200\mathrm{c}}}{M_0}\right)\right], \label{eq:sigmalnMwl}
\end{equation}
with a pivot mass $M_0 = 2\times10^{14}\,h^{-1}\Msun$.

We create synthetic cluster shear maps by applying the source redshift distribution, cluster miscentering, and cluster member contamination to halo mass maps from numerical simulations, and we then use these maps to calibrate the free parameters of the \Mwl--\Mhalo\ relation \citep{grandis21}.\footnote{All systematic and stochastic uncertainties in the lensing model and data, in halo morphology, and in the projected large-scale structure along the cluster line of sight are thus accounted for in our \Mwl--\Mhalo\ relation.}
Note that we use the mass maps from full-physics hydrodynamical simulations and establish the relationship between \Mwl\ and the ``gravity-only mass'' \Mhalo\ as measured in paired gravity-only simulations with identical initial conditions.
This has the benefit that our model includes the effect of baryons, but we are still able to use the arguably more robust halo mass function predictions from gravity-only simulations \citep[such as][]{tinker08} (including emulators e.g., \cite{mcclintock19emu, nishimichi19, bocquet20}).
In practice, we apply this method to the Magneticum \citep{hirschmann14, teklu15, beck16, dolag17} and the Illustris TNG simulations \citep{pillepich18, marinacci18, springel18, nelson18, naiman18, nelson19}.
The recovered constraints on the \Mwl--\Mhalo\ relation differ somewhat, which we interpret as an uncertainty in the modeling of baryonic effects.
We inflate the uncertainty on all parameters of the \Mwl--\Mhalo\ relation by this additional uncertainty given in Table~2~\cite{grandis21} that, e.g., for the amplitude $b_\mathrm{WL}$, amounts to 2\%.
As we will show below, this level of systematic uncertainty is smaller than the current level of statistical uncertainty in the lensing dataset, and our analysis therefore does not strongly depend on the assumption that the Magneticum or Illustris TNG simulation correctly reproduces the real universe.

For the HST-39 dataset, we use a similar modeling approach, and also adopt \Mwl\ as a latent variable.
Note, however, that several analysis choices differ.
For example, the selection of HST lensing sources via color cuts leads to a much smaller level of cluster member contamination than in the DES~Y3 analysis. Therefore, we do not explicitly model this contamination, and we do not explicitly model the effect of miscentering, either.
We do, of course, take all sources of uncertainty into account in the \Mwl--\Mhalo.
The full HST cluster lensing model is presented in detail in the original analyses \cite{schrabback18, schrabback21, zohren22}.

\subsection{Likelihood, Pipeline, Priors, and Sampling}

We infer parameters $\vec p$ from our data assuming a cluster population model.
Bayes' theorem states
\begin{equation}
  P(\vec p|\mathrm{data}, \mathrm{model}) \propto P(\mathrm{data}|\vec p, \mathrm{model})\, P(\vec p| \mathrm{model}),
\end{equation}
where $P(\vec p| \mathrm{model})$ is the prior probability distribution of the parameters.
The factor $P(\mathrm{data}|\vec p, \mathrm{model})$ is the likelihood $\mathcal L$ of the data given the (fixed) model and the parameters.
A detailed discussion of our likelihood function and its numerical implementation can be found in \citetalias{bocquet24I}, Sec.~VII.
Briefly, we adopt a log-likelihood
\begin{equation}
  \label{eq:likelihood}
  \begin{split}
    \ln \mathcal L =& \ln \mathrm{Poisson}\left(\text{sample}\left\{\xi_i,\lambda_i,z_i\right\}_{i=1}^{N_\text{clusters}} \bigg| \frac{\dif^3 N(\vec p)}{\mathop{\dif\xi} \mathop{\dif\lambda} \mathop{\dif z}}\right) \\
    &+ \sum_i \ln P\left(\lambda_i,\, \vec g_{\mathrm{t},i} | \lambda_i>\lambda_\mathrm{min}(z), \xi_i, z_i, \vec p\right),
  \end{split}
\end{equation}
where the first term is the unbinned Poisson likelihood of the cluster sample $\left\{\xi,\lambda,z\right\}$ and its distribution $\frac{\dif^3 N(\boldsymbol p)}{\mathop{\dif\xi} \mathop{\dif\lambda} \mathop{\dif z}}$ in $\xi$--$\lambda$--$z$ space, and the second term is the ``mass calibration likelihood'': the conditional probability to observe the tangential shear profile $\vec g_\mathrm{t}$ and richness $\lambda$ for a given cluster in the sample with measured $\xi$ and $z$, and whose richness is above the minimum richness $\lambda_\mathrm{min}$. 

We implement the likelihood function as a Python module in the \textsc{CosmoSIS} framework \citep{zuntz15}.\footnote{\url{https://cosmosis.readthedocs.io/}}
We validate the implementation of the analysis pipeline by analyzing several statistically independent mock catalogs that we create from the model.
Because creating those mocks is less challenging than analyzing them, this is a meaningful test of the analysis pipeline (rather than a test of the mock creation).
For further details, we refer the reader to \citetalias{bocquet24I}, Sec.~VIII.

\begin{table}
  \caption{Fit parameters in the baseline analysis of the abundance of SPT clusters with DES~Y3 and HST-39 lensing.
  A missing entry (-) for the prior indicates that a uniform prior is applied that is wide enough to not be informative.
  Priors with a lower limit ($>$) are uniform above that bound with an un-informative upper bound.
  The parameters of the lensing models are prior-dominated and inform the empirical calibration of the other observable--mass relations.
  The prior on \Onuhh\ corresponds to a prior on the sum of neutrino masses $\sumMnu\sim\mathcal U(0, 0.6)~\mathrm{eV}$.
  \label{tab:parameters}}
  \begin{ruledtabular}
    \begin{tabular}{lll}
      Parameter & Description & Informative Prior\\
      \colrule
      \multicolumn{3}{l}{DES Y3 cluster lensing} \\
      $\sigma_{\ln b_\mathrm{WL,1}}$ & scaling of  bias & $\mathcal N(0, 1)$ \\
      $\sigma_{\ln b_\mathrm{WL,2}}$ & scaling of bias & $\mathcal N(0, 1)$ \\
      $b_{\mathrm{WL},M}$ & mass slope of bias & $\mathcal N(1.029, 0.006^2)$ \\
      $s_\mathrm{WL}$ & scaling of scatter & $\mathcal N(0, 1)$ \\
      $s_{\mathrm{WL},M}$ & mass slope of scatter & $\mathcal N(-0.226,0.040^2)$ \\
      \colrule
      \multicolumn{3}{l}{HST cluster lensing} \\
      $b_\mathrm{HST}$ & overall scaling of bias & $\mathcal N(0, 1)$ \\
      $s_\mathrm{HST}$ & overall scaling of scatter & $\mathcal N(0, 1)$ \\
      \colrule
      \multicolumn{2}{l}{SZ--mass parameters} \\
      $\ln\asz$ & amplitude & \dots\\
      \bsz & mass slope & \dots\\
      \csz & redshift evolution & \dots\\
      \sigmalnzeta & intrinsic scatter & $>0.05$\\
      $\gamma_\mathrm{ECS}$ & depth of SPTpol ECS & \dots\\
      \colrule
      \multicolumn{3}{l}{DES richness--mass parameters (used for $z<1.1$)} \\
      $\ln\alambda$ & amplitude & \dots\\
      \blambda & mass slope & \dots\\
      \clambda & redshift evolution & \dots\\
      \sigmalnlambda & intrinsic scatter & $>0.05$\\
      \colrule
      \multicolumn{3}{l}{WISE richness--mass parameters (used for $z>1.1$)} \\
      $\ln\alambda$ & amplitude & \dots\\
      \blambda & mass slope & \dots\\
      \clambda & redshift evolution & \dots\\
      \sigmalnlambda & intrinsic scatter & $>0.05$\\
      \colrule
      \multicolumn{3}{l}{Correlation coefficients} \\
      $\rho_\mathrm{SZ,WL}$ & SZ--weak-lensing & $\mathcal U(-0.5, 0.5)$ \\
      $\rho_\mathrm{SZ,\tilde\lambda}$ & SZ--richness & $\mathcal U(-0.5, 0.5)$ \\
      $\rho_\mathrm{WL,\tilde\lambda}$ & weak-lensing--richness & $\mathcal U(-0.5, 0.5)$ \\
      \colrule
      \multicolumn{3}{l}{Cosmology} \\
      \Om\ & matter density & $>0.1$ \\
      \Onuhh\ & neutrino density & $\mathcal U(0, 0.00644)$ \\
      \Obhh & baryon density & $\mathcal N(0.02236,0.00015^2)$ \\
      $h$ & Hubble parameter & $\mathcal N(0.7, 0.05^2)$ \\
      $\ln10^{10}A_s$ & amplitude of $P(k)$ & \dots \\
      $n_s$ & scalar spectral index & $\mathcal N(0.9649, 0.0044^2)$ \\
      $w$ & dark energy equation & $-1$ (\LCDM)\\
       & ~~~of state parameter & $\mathcal U(-2, -0.4)$ (\wCDM)
    \end{tabular}
  \end{ruledtabular}
\end{table}

The parameters that we vary in the baseline analysis are defined in Table~\ref{tab:parameters}, along with the informative priors where applied.
The model for the DES cluster lensing mass bias and scatter are functions of cluster redshift, and the parameters $\sigma_{\ln b_\mathrm{WL,1/2}}$ and $s_\mathrm{WL}$ modulate the uncertainty of the model (therefore, the priors are centered on 0 with unit width) \citepalias[see][Sec.~V~B]{bocquet24I}.
We follow the same approach for the HST lensing data, where each cluster has its own bias and scatter, and the shared uncertainties are modulated via $b_\mathrm{HST}$ and $s_\mathrm{HST}$.
The priors on the scatter ($>0.05$) improve the numerical stability of the code.
The priors on the correlation coefficients $\mathcal U(-0.5,0.5)$ are chosen such that the resulting correlation matrix is nonsingular for all parameter combinations.
The parameters $h$, \Obhh, and $n_s$ are not constrained by the cluster dataset, and are not significantly correlated with the parameters of interest.
We apply Gaussian priors on \Obhh\ and $n_s$ from {\it Planck} \citep{planck18VI}, and a prior on the Hubble parameter $h\sim\mathcal N(0.70, 0.05^2)$ that is wide enough to encompass all currently favored constraints.
When analyzing the cluster dataset jointly with {\it Planck}, we do not apply these three Gaussian priors, and we additionally consider the optical depth to reionization $\tau$ as a free parameter.

We explore the parameter space using either the \textsc{multinest} or the \textsc{nautilus} sampler \citep{feroz09multinest, lange23nautilus}.
The potential amount by which these samplers may or may not underestimate the parameter uncertainties is small and does not change our findings in any qualitatively relevant way \citep{lemos23}.

\section{Blinding, Robustness Tests, and Unblinding}
\label{sec:robustness_tests}

In this section, we describe our blinding strategy and the robustness tests that were performed prior to unblinding of the results.
We check for internal consistency of the dataset and verify the robustness to several analysis choices in the processing of the DES~Y3 lensing data.

\subsection{Blinding Strategy}

To avoid compromising our analysis with any expectations we may have regarding what the results should or should not look like, we proceed as follows.
First, the SPT cluster catalogs and the DES Y3 lensing data were finalized before we started this analysis.
Second, we did not compare any weak-lensing mass estimates or preliminary SZ--mass relations with literature values.
Third, we blinded the astrophysical and cosmological constraints obtained in this work until the robustness of the results against specific analysis choices was demonstrated.
To do so, we created unknown, random, but constant \emph{blinding offset parameters} with which we shifted the parameters when reading in any MCMC chain.
It was thus possible to compare different blinded analyses with each other to check for relative shifts in the recovered parameters (or changes in the parameter uncertainties) without knowing the actual parameter values.

The HST-39 cluster lensing dataset was finalized, and its analysis published before we started this analysis.
Therefore, the parameter constraints favored by this dataset were known, and we preferred to not use these data in the blinded phase of the analysis.

\begin{figure*}
  \includegraphics[width=\textwidth]{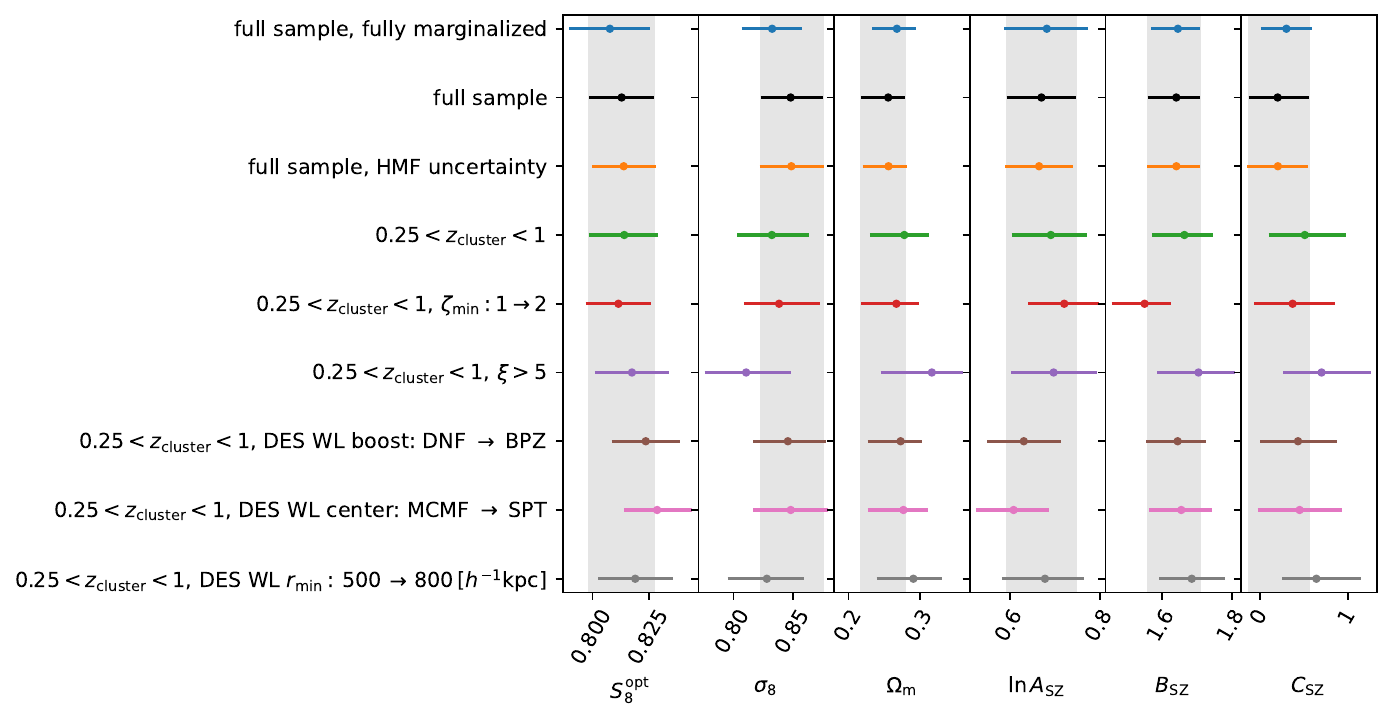}
  \caption{Impact of different analysis choices on the constraints on key cosmological and SZ--mass relation parameters (mean and 68\% credible interval), assuming a \LCDM\ model.
  During the blinded phase of the analysis, the parameter values were artificially offset by an unknown amount.
  The vertical gray bands show the 68\% credible interval of the ``full sample'' analysis (second row) for reference.
  All results shown here are based on mass calibration using only the DES~Y3 lensing data, the HST-39 lensing data are not used.
  Only the results in the first row shows constraints that are marginalized over all nuisance parameters; for simplicity, for all other tests, we fix the parameters of the \Mwl--\Mhalo\ relations, the intrinsic scatter correlation coefficients, and $\Omega_{\nu}h^2$, $\Omega_{b}h^2$, and $n_s$ to their nominal values.
  DNF and BPZ are two redshift estimators that we use to determine the amount of cluster member contamination in the lensing signal.
  MCMF and SPT refer to cluster centers as measured in optical data by MCMF or as determined in the SZ analysis of SPT data.
  }
  \label{fig:robustness}
\end{figure*}

The blinded runs were set up assuming a \LCDM\ model with wide flat priors on the fit parameters.
Failure to choose sufficiently wide priors might lead to the blinded results showing hard cuts, thereby effectively unblinding the respective parameter constraint.
Note that arbitrarily wide priors are not always possible, because, e.g., the scatter parameters or \Om\ need to be strictly positive.
All but the first test are summarized in Fig.~\ref{fig:robustness}, which shows the constraints on key parameters.
The results in the first row (blue) are obtained from the full dataset (without HST-39), marginalized over all parameters in Table~\ref{tab:parameters}.
The second row (black) shows the constraints from the same analysis, except that we fix a selection of model parameters to their fiducial values: the parameters of the \Mwl--\Mhalo\ relations, the correlation coefficients on intrinsic scatter, \Obhh, \Onuhh, and $n_s$.
The parameter shifts are small, which also suggests that our constraints are not limited by the systematic uncertainty in the DES~Y3 lensing model.
To save computational resources, and to make the robustness tests somewhat more stringent, all cross-check analyses discussed in the next section are run in this restricted parameter space, with a selection of parameters fixed as just discussed.

Before running the blinded robustness tests, we had agreed that we would unblind the results if no test reveals a statistically significant difference with the baseline analysis.

\subsection{Tests Performed prior to Unblinding}
We now summarize the tests that were conducted with the blinding setup discussed above.

\subsubsection{Combination of Mass Calibration and Abundance}
\begin{figure*}
  \includegraphics[width=\textwidth]{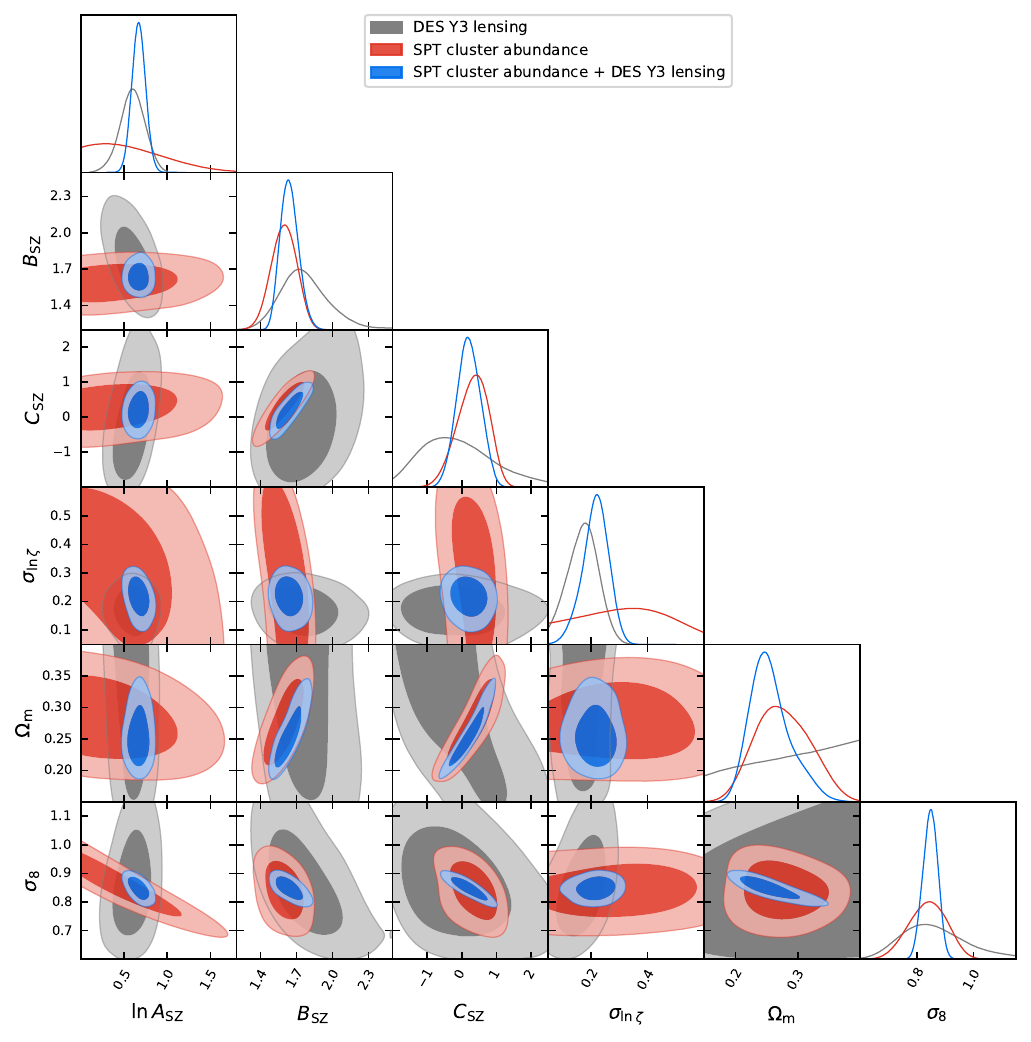}
  \caption{Constraints (68\% and 95\% credible regions) on select scaling relation and cosmology parameters, as obtained from just the DES Y3 weak-lensing-based mass calibration (gray) or just the cluster abundance (red).
  The abundance and mass calibration results do not show signs of significant disagreement, and we combine them to obtain cosmological constraints from the joint analysis of cluster abundance and lensing (blue).
  }
  \label{fig:combinations}
\end{figure*}

We derive constraints from just the cluster abundance [first term in the likelihood Eq.~(\ref{eq:likelihood})] and just mass calibration (second term in the equation).
We do not expect interesting constraints on the cosmological parameters from these individual analyses, and Fig.~\ref{fig:combinations} shows broad uncertainties in the cosmological parameters and elongated parameter degeneracies (gray and red contours).\footnote{The mass calibration analysis constrains the parameters of the observable--mass relations but does place interesting constraints on the cosmological parameters.
The analysis of the abundance measurement cannot tightly constrain the cosmological parameters, either, because these are degenerate with the scaling relation parameters (this is simply a manifestation of the fact that cluster abundance cosmology requires mass information).}
The two posterior distributions overlap, and we combine the two analyses to obtain constraints from the cluster abundance with weak-lensing-informed mass calibration (blue contours in Fig.~\ref{fig:combinations}).
The constraints on select parameters are shown labeled as ``full sample'' in Fig.~\ref{fig:robustness}.
We note that the situation is similar for the parameters of the richness--mass relations, for which the individual constraints overlap, and tighten when we combine the abundance with the lensing mass calibration.

\subsubsection{Goodness of Fit}
We validate that the model is an adequate description of the data by extracting posterior predictive distributions from the blinded MCMC chains (of the cluster abundance and lensing analysis) and comparing them with the data.
In Fig.~\ref{fig:dN_dzdxidlambda}, blue lines show the mean predicted model for the cluster abundance.
Similarly, in Fig.~\ref{fig:stacked_shear}, we show the mean predicted stacked shear profiles and the error on the mean.
We compute the statistical difference between each stack and the mean model, and obtain values for $\chi^2\in\{11.3,8.1, 7.3, 10.6\}$ for $\{7, 7, 6, 6\}$ data points in the $\{(z<0.5,\,\xi\geq5.6),\,(z<0.5,\,\xi<5.6),\,(z\geq0.5,\,\xi\geq5.6),\,(z\geq0.5,\,\xi<5.6)\}$ bins.
For the full dataset, we measure $\chi^2=37.4$ for 26 data points, and conclude that the model adequately describes the data.

\subsubsection{Uncertainty in the Halo Mass Function}
We marginalize over a 3.5\% uncertainty in the amplitude and a 2\% uncertainty in the slope of the mass function (following \cite{DESY1cl}).
The lack of parameter shifts in Fig.~\ref{fig:robustness} suggests that we can neglect uncertainties in the halo mass function, which we will do from here on out.\footnote{We note that this test did not necessarily need to be run before unblinding.}

\subsubsection{Restricting the Analysis to Low Redshift $z<1$}

At redshifts beyond $z\gtrsim1$, it remains unclear how strong the correlated astrophysical emission, and thus the contamination of the cluster SZ signal is, although observational studies favor an effect that is only at the few-percent level e.g., \cite{gupta17, melin18, zubeldia23, hashiguchi23}.
Also recall that we use DES data for cluster confirmation up to redshift $z=1.1$, beyond which we use data from WISE.
When limiting the analysis to $z<1$, we thus test the robustness of using WISE data for cluster confirmation and redshift assignment jointly with the robustness against astrophysical contamination of the SZ signal.
We caution that disagreement between the fiducial sample and the low-redshift subset could also indicate interesting new physics (e.g., different growth of structure beyond $z\gtrsim1$) rather than pointing to the effects we might worry about.

The constraints from the $0.25<z_\mathrm{cluster}<1$ dataset (869 clusters compared to 1,005 in the full sample) show negligible shifts compared to the analysis of the full sample (see Fig.~\ref{fig:robustness}).
Since the remaining robustness checks are not expected to be sensitive to the redshift range of the sample, we perform the following tests on the $0.25<z_\mathrm{cluster}<1$ sample.

\subsubsection{The $\xi-\zeta$ Relation}
The $\xi-\zeta$ relation [Eq.~(\ref{eq:xizeta})] cannot hold down to arbitrarily low significance as the formalism breaks down for $\xi<\sqrt{3}$.
As in previous work, we define a limit $\zeta_\mathrm{min}$, below which the maximization bias in the $\xi-\zeta$ relation does not apply.
Using simulations of the SZ sky, we estimate $\zeta_\mathrm{min}=1$.
Here, we confirm that raising the threshold to $\zeta_\mathrm{min}=2$ (which was applied in, e.g., \citep{dehaan16}) only has a small impact on the recovered parameters (``$\zeta_\mathrm{min}:\,1\rightarrow2$'' in Fig.~\ref{fig:robustness}).

\subsubsection{Selection Cut in SPT Detection Significance $\xi$}
Enabled by the DES data and the MCMF analysis, in this study, we select clusters down to lower detection significances than previously done [see Eq.~(\ref{eq:select_DES})].
We compare the constraints obtained using this selection scheme (869 clusters with $z<1$)
to the constraints obtained using the more conservative $\xi>5$ selection (654 clusters with $z<1$).
As shown in Fig.~\ref{fig:robustness} as ``$\xi>5$'' in purple, the constraint on \Sopt\ is robust to this change, but the SZ mass slope \bsz\ and the redshift evolution \csz\ both shift to higher values, and, due to parameter degeneracies, \Om\ also shifts to larger values.

We quantify these shifts more carefully.
Since \Sopt\ is effectively unchanged, and because of the parameter degeneracies, it is sufficient to focus on the constraints on \Om.
First, we consider the analysis of the ``low-redshift'' $0.25<z_\mathrm{cluster}<1$ dataset (green in Fig.~\ref{fig:robustness}), and examine the impact of additionally applying $\xi>5$.
Naively, one would compare the difference in the mean recovered values for \Om\ with the expected variance, computed as the sum of the variances from each analysis.
Doing so, we observe a shift in \Om\ that is at the $0.5\sigma$ level.
However, this simple estimate neglects the fact that the $\xi>5$ sample is a subset of the larger catalog.
In this case, ``the covariance of the parameter differences between the partial and full analyses is simply the difference of the covariances'' \citep{gratton&challinor20}.
With this estimator, we report that \Om\ shifts at the $0.9\sigma$ level.

To empirically validate that this level of parameter shift is expected, we analyze pairs of mock catalogs (restricted to $z<1$)
for which one catalog is additionally restricted to $\xi>5$.\footnote{We use the mock catalogs created for validating the analysis pipeline in \citetalias{bocquet24I}.}
After running two such mock analyses, we do indeed observe parameter shifts that are very similar to the one obtained from the real data.

Finally, we compute the consistency of the \Om\ measurement between the full cluster sample ($z>0.25$) and the $\xi>5, 0.25<z<1$ subsample (purple vs. black in Fig.~\ref{fig:robustness}).
Naively adding the variances gives a shift at the $0.9\sigma$ level, and correctly subtracting the variances yields $1.2\sigma$.
We caution the reader that in this comparison, we consider the combined effects of restricting ourselves to low redshifts (which shifts \Om\ toward higher values) and to high significances (which further shifts \Om\ high).
To summarize, we find that, while the observed shifts are interesting and might reveal new properties of SZ-selected clusters in upcoming analyses of larger samples, there is no statistically significant evidence of internal tension in the dataset.

\subsubsection{Cluster Member Contamination}
In the DES~Y3 lensing analysis, we account for cluster member contamination of the lensing source catalog as determined by applying the $P(z)$ decomposition method to DNF source redshifts (see \citetalias{bocquet24I}, Sec.~IV~D).
However, the level of contamination derived using BPZ redshifts is significantly different.
We test the robustness of our cosmological results to adopting the cluster member contamination as derived using BPZ instead of the fiducial one derived using DNF.
The resulting parameter shifts are $\lesssim1\sigma$ (see ``DES WL boost: DNF~$\rightarrow$~BPZ'' in Fig.~\ref{fig:robustness}).
While this level of shifts is acceptable given the overall level of uncertainty, this test highlights the relative importance of the cluster member contamination in our analysis.
Future lensing surveys are expected to enable a cleaner source selection, e.g., \cite{Euclid24XXXVII}, which would reduce that relative importance.

\subsubsection{Centers Adopted for the Lensing Analysis}
Our default analysis uses lensing data extracted around cluster positions as determined using the DES imaging data.
We constrain the optical miscentering distribution and account for it in the lensing analysis (see \citetalias{bocquet24I}, Secs. IV~C and V~B).
As an alternative, we use the cluster centers as determined in the SZ analysis, and account for the corresponding miscentering distribution in the lensing model.
The recovered constraints on \asz\ and \Sopt\ shift by $\lesssim1\sigma$, while the other parameters remain essentially unchanged (``DES WL center: MCMF~$\rightarrow$~SPT'' in Fig.~\ref{fig:robustness}).
We conclude that our miscentering models and corrections are adequate.

\subsubsection{Radial Range in the Lensing Analysis}
Our fiducial setup ($r>500\,h^{-1}\mathrm{kpc}$) is a compromise between avoiding the cluster central regions (where miscentering, cluster member contamination, baryonic feedback, and possible non-linear shear responses are particularly challenging to model), and including scales with high weak-lensing signal-to-noise ratios \citep{grandis21}.
We validate this choice by comparing the fiducial results to an analysis with an even more conservative radial cut, $r>800\,h^{-1}\mathrm{kpc}$.
The outer limit remains unchanged: $r<3.2\,(1+z_\mathrm{cluster})^{-1}\,h^{-1}\mathrm{Mpc}$.
As expected, the recovered parameter constraints slightly broaden compared to the fiducial setup, but the central values remain almost identical (last row in Fig.~\ref{fig:robustness}, ``DES WL $r_\mathrm{min}$:~500~$\rightarrow~800\,h^{-1}\mathrm{Mpc}$).
This confirms that our model can adequately describe cluster shear profiles down to $r=500\,h^{-1}\mathrm{kpc}$.

\subsection{Unblinding}
The validation tests discussed in the previous section were designed to probe what we expected to be the most probable points of failure of our model.
We performed these tests while being blinded to the actual values of the mean recovered parameter constraints.
Since no alternative analysis led to significant parameter shifts, we proceeded and removed the so-far unknown blinding parameter offsets.
Throughout this work, we present the results as obtained after unblinding, with no further modification of the data or the analysis.

\section{Results}
\label{sec:results}

We present and discuss constraints on the \LCDM\ and \wCDM\ models, and on the scaling relation parameters.
Our baseline setup includes the sum of neutrino masses \sumMnu\ as a free parameter; however, we can only constrain it in combination with measurements of primary CMB anisotropies from {\it Planck}.
We also present constraints on an empirical extension of the \LCDM\ model, in which we allow the rate of structure growth to deviate from the standard prediction.
We implement this test as a direct measurements of \sig\ at different redshifts between 0.25 and 1.8.

\begin{figure}
  \includegraphics[width=\columnwidth]{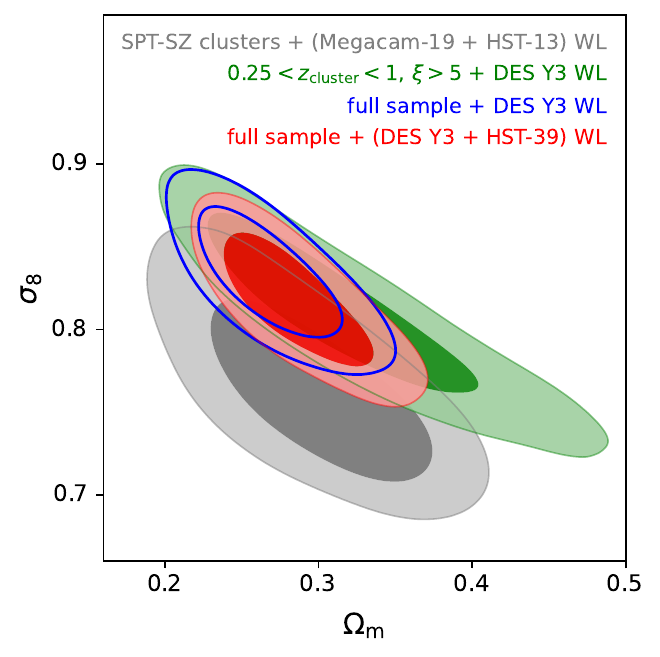}
  \caption{Constraints on \Om\ and \sig\ from the abundance of SPT-selected clusters (68\% and 95\% credible regions).
  In comparison with the precursor analysis of 343~SPT-SZ clusters (gray contours; \cite{bocquet19}), the larger cluster sample and the improved weak-lensing mass calibration enable tighter constraints (red contours).
  We also show the constraints obtained from the SPT cluster abundance with DES Y3 lensing, without using the HST-39 high-redshift cluster lensing dataset (blue contours).
  Green contours show the constraints obtained when restricting the cluster sample to $\xi>5$, as was done in precursor analyses (this result also corresponds to item~6 in Sec.~\ref{sec:robustness_tests}).
  }
  \label{fig:Om-sig8_SPTcl}
\end{figure}

\subsection{\LCDM\ with Massive Neutrinos}
\label{sec:LCDM}

In Fig.~\ref{fig:Om-sig8_SPTcl}, we show the constraints on \Om\ and \sig\ as derived from the SPT cluster abundance.
We note that supplementing the DES lensing data with high-redshift cluster lensing data from HST leads to a slight shift of the constraints along the degeneracy axis.
The results obtained from the fiducial, combined analysis setup---SPT(SZ+pol) cluster abundance with (DES Y3 + HST-39) lensing---are summarized in Table~\ref{tab:results}.
As demonstrated in Fig.~\ref{fig:Om-sig8_SPTcl}, our new analysis significantly improves over the precursor study, which used 343~clusters in the SPT-SZ survey and lensing information from 32 targeted observations \citep{bocquet19}.
Our new results are shifted toward higher values of \sig, which is caused by using DES~Y3 lensing data for mass calibration.
Remember that our analysis also uses a larger cluster sample than the predecessor analysis, but the SPT cluster samples are consistent with one another, ruling out the possibility that the shift in \sig\ can be explained with changes in the cluster sample \citep{bleem20, bleem24}.\footnote{We checked that analyzing the SPT cluster sample used in this work with the lensing data used in \cite{bocquet19} leads to results that are centered on the gray contours in Fig.~\ref{fig:Om-sig8_SPTcl}, whereas analying the 343 SPT-SZ clusters with the DES~Y3 + HST-39 lensing data leads to constraints centered on the results presented in this work.}
Finally, we note that neglecting the systematic uncertainties in the lensing model only has a small impact on the recovered constraints (first vs. second row in Fig.~\ref{fig:robustness}).
This implies that our constraints are still limited by the precision of the lensing measurements, rather than the uncertainties in the modeling.
The statistical uncertainty will be improved in future analyses by increasing the number of lenses (larger cluster sample), and/or by using lensing datasets with higher source densities and higher lensing efficiencies.

\begin{figure}
  \includegraphics[width=\columnwidth]{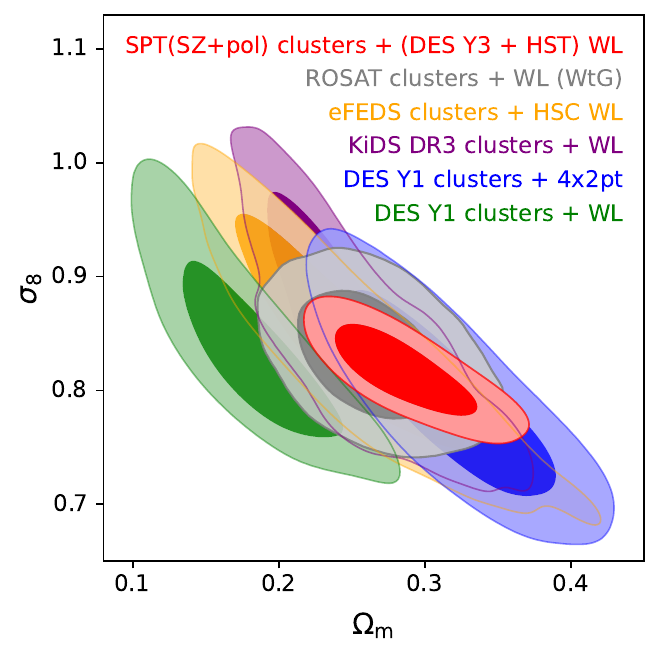}
  \caption{Constraints on \Om\ and \sig\ from various analyses of the cluster abundance (68\% and 95\% credible regions).
  Along with our work, we also show results obtained from X-ray selected clusters (ROSAT \citep{mantz15}, eFEDS \citep{chiu23}) and optically selected clusters (DES~Y1 \citep{DESY1cl, to21results}, KiDS \citep{lesci22}).
  Our analysis places the tightest constraints on \sig\ and \Sopt.
  }
  \label{fig:Om-sig8_cl}
\end{figure}

\begin{table*}
  \caption{Parameter constraints for the \LCDM\ and \wCDM\ models (mean and 68\% credible intervals, or 95\% limit).
  The SPTclusters+lensing dataset corresponds to SPT(SZ+pol) clusters + (DES Y3 + HST-39) WL.
  The parameters $n_s$ and \Obhh\ are not meaningfully constrained by the cluster dataset and the results are thus omitted here.
  Similarly, the cluster dataset cannot constrain $h$ and \sumMnu\ on its own, and we only quote the results recovered in combination with {\it Planck} 2018 TT,TE,EE+lowE \citep{planck18VI}.
  \label{tab:results}}
  \begin{ruledtabular}
    \begin{tabular}{lcccccccc}
      Dataset & \Om & \sig & $S_8$ & \Sopt\ & $h$ & \sumMnu~[eV] & $w$\\
      \colrule
      \LCDM\\
      SPTclusters+lensing & $0.286\pm0.032$ & $0.817\pm0.026$ & $0.795\pm0.029$ & $0.805\pm0.016$ & \dots & \dots & $-1$\\
      SPTclusters+lensing+{\it Planck} & $0.315\pm0.011$ & $0.807\pm0.013$ & $0.827\pm0.013$ & $0.817\pm0.013$ & $0.674\pm0.008$ & $<0.18$ & $-1$\\
      \colrule
      \wCDM\\
      SPTclusters+lensing & $0.268\pm0.037$ & $0.820\pm0.026$ & $0.772\pm0.040$ & $0.796\pm0.020$ & \dots & \dots & $-1.45\pm0.31$\\
      SPTclusters+lensing+{\it Planck} & $0.257\pm0.026$ & $0.848\pm0.027$ & $0.783\pm0.026$ & $0.814\pm0.016$ & $0.75\pm0.04$ & $<0.6$ & $-1.34^{+0.22}_{-0.15}$\\
    \end{tabular}
  \end{ruledtabular}
\end{table*}

We compare our results with other analyses of the cluster abundance and lensing mass calibration (see Fig.~\ref{fig:Om-sig8_cl}).
The results from the DES Y1 cluster abundance and lensing are shifted toward lower values of \Om\ than expected from most cosmological probes \citep{DESY1cl}, and are wider than the constraints we report here.
The same cluster sample, cross-correlated with clusters, galaxies, and shear on large scales (``DES~Y1 clusters + 4$\times$2~pt''), yields constraints that largely overlap with ours \citep{to21results}.
The analyses of optically selected clusters in the KiDS survey with KiDS lensing \citep{lesci22}, and the analysis of clusters selected in the eROSITA Science Verification data (eFEDS) with HSC lensing \citep{liu22eFEDS, chiu23} both exhibit constraints that are less tightly constrained along the direction of the \Om--\sig\ degeneracy than our analysis, but they overlap with our results.
The result obtained from ROSAT clusters with Subaru/HSC lensing ``Weighing the Giants'' (WtG) is almost centered on ours, but is somewhat less constraining along the width of the degeneracy \citep{mantz15}.
We note that their constraint also includes a constraint on \Om\ through measurements of gas mass fractions $f_\mathrm{gas}$, whereas our constraints are solely derived from the cluster abundance.

\begin{figure}
  \includegraphics[width=\columnwidth]{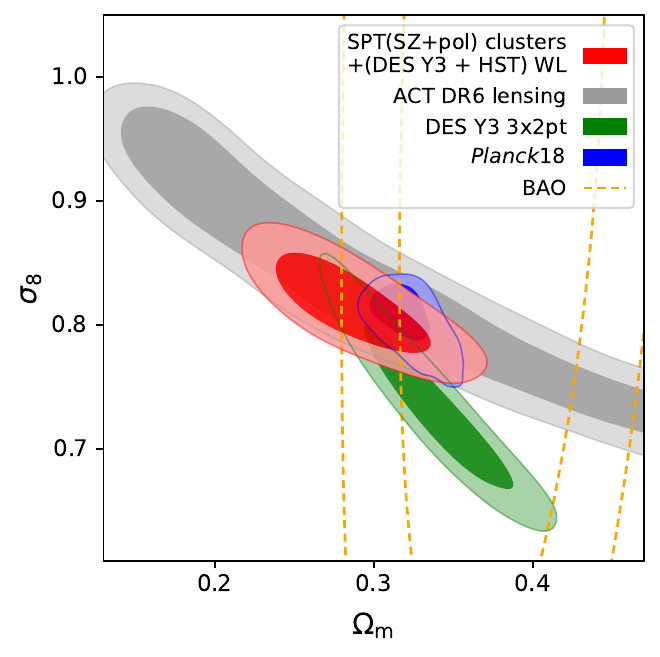}
  \caption{Constraints on \Om\ and \sig\ (68\% and 95\% credible regions) in a \LCDM\ universe with massive neutrinos from the abundance of SPT clusters, a selection of large-scale structure lensing probes \citep{DESY33x2pt, qu24}, {\it Planck} primary CMB measurements (TT,TE,EE+lowE) \citep{planck18VI}, and BAO \citep{beutler11, ross15, alam17, alam21}.
  Our measurement of the cluster abundance places competitive constraints especially on \Sopt.
  }
  \label{fig:Om-sig8}
\end{figure}

In Fig.~\ref{fig:Om-sig8}, we compare our constraints in the \Om--\sig\ plane with those obtained from other cosmological probes.
Current measurements on CMB lensing from ACT and SPT cannot break the \Om--\sig\ parameter degeneracy, but their degeneracy goes through our constraints \citep{bianchini20, madhavacheril24, pan23SPT3Glensing}.
Our measurement $S_8^\mathrm{opt}=0.805\pm0.016$ and $S_8^\mathrm{opt}= 0.818\pm0.022$ from ACT DR6 CMB lensing alone \citep{qu24} agree to within $0.5\sigma$.
As an example of joint analyses of weak lensing and galaxy clustering, we show the DES Y3 3$\times$2~pt results \citep{DESY33x2pt}.
They constrain $S_8^\mathrm{opt} = 0.754\pm0.031$, which is $1.5\sigma$ lower than our measurement.
In the \Om--\sig\ plane, the difference between their results and ours has a probability to exceed of 0.25 ($1.1\sigma$).\footnote{We compute the probability to exceed using \url{https://github.com/SebastianBocquet/PosteriorAgreement} (see Sec.~5.2 in \cite{bocquet15}).}
Finally, {\it Planck} 2018 TT,TE,EE+lowE data (without lensing) constrain $S_8^\mathrm{opt} = 0.820\pm0.016$ \citep{planck18VI}, which differs from our measurement by $0.7\sigma$, and with an uncertainty that is equal to ours.
In the \Om--\sig\ plane, the difference is essentially identical (probability to exceed of $0.47,\,0.7\sigma$).
For the $S_8$ parameter, the difference is $1.1\sigma$, which does not strengthen the claim of an ``$S_8$ tension'' between large-scale structure measurements at late times and the primary CMB.
We also show constraints from baryonic acoustic oscillations (BAO), which constrain a combination of \Om\ and $H_0$, but are unable to constrain \sig.\footnote{Specifically, we analyze BAOs as measured using the 6dFGS \citep{beutler11}, SDSS DR7 Main Galaxy Sample \citep[MGS,][]{ross15}, BOSS DR12 luminous red galaxies \citep[LRGs,][]{alam17}, and eBOSS DR16 LRGs \citep{alam21} datasets.}

\begin{figure}
  \includegraphics[width=\columnwidth]{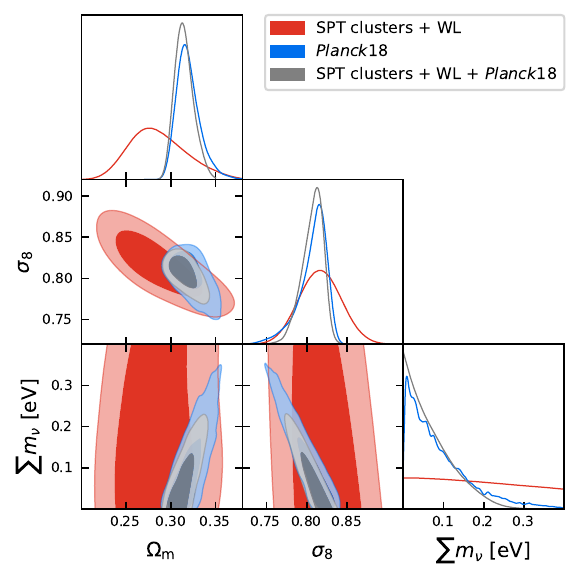}
  \caption{Constraints (68\% and 95\% credible regions) on \Om, \sig, and on the sum of neutrino masses \sumMnu\ from the abundance of SPT clusters and its combination with {\it Planck} primary CMB anisotropies (TT,TE,EE+lowE) \citep{planck18VI}.
  The joint analysis places an upper limit $\sumMnu<0.18~\mathrm{eV}$ (95\%~limit).
  }
  \label{fig:cl_planck_nuLCDM}
\end{figure}

We proceed and combine our cluster abundance measurements with primary CMB anisotropies ({\it Planck} 2018 TT,TE,EE+lowE).
As shown in Fig.~\ref{fig:cl_planck_nuLCDM}, the {\it Planck} analysis exhibits a well-known degeneracy between the sum of neutrino masses \sumMnu, \Om, and \sig.
Conversely, the cluster abundance---like other probes of the large-scale structure---constrains \sig\ essentially independently of \sumMnu.
The combined analysis therefore leads to improved constraints by breaking that degeneracy.
Our measurement of \sig\ fully encompasses the {\it Planck} measurement, and the cluster measurements only help to exclude the high-\sumMnu\ tail.
We report an improved constraint on the upper limit,
\begin{equation}
    \sumMnu<0.18~\mathrm{eV}~ (95\%~ \mathrm{limit}),
\end{equation}
compared to $\sumMnu<0.26~\mathrm{eV}$ from {\it Planck} data alone.
The constraints on the other cosmological parameters only improve mildly (see Table~\ref{tab:results}).

\subsection{Constraints on the Observable--Mass Relations}

In our analysis, we simultaneously calibrate the observable--mass relations while fitting for cosmology.
In Table~\ref{tab:scaling_results}, we present the recovered constraints on the parameters of the relations defined in Eqs.~(\ref{eq:zetaM}) and~(\ref{eq:lambdaM}).
In the Appendix, we show marginalized parameter constraints for all scaling relation and cosmological parameters that are not prior-dominated.

\subsubsection{The SZ--Mass Relation}
The SZ parameter constraints are similar to, but tighter than the ones presented in the precursor SPT cluster analysis based on targeted lensing observations \citep{bocquet19}, with the exception of the amplitude \asz.\footnote{Note that we are using $M_{200\mathrm{c}}$ in this work, as opposed to $M_{500\mathrm{c}}$ in \citep{bocquet19}. Therefore, the recovered constraints on \asz\ can only be compared after converting between the two mass definitions.}
Our new amplitude is somewhat lower than the old result, which also explains the shift toward larger \sig\ seen in Fig.~\ref{fig:Om-sig8_SPTcl} (a lower \asz\ implies that the observed clusters are more massive).
As discussed in Sec.~\ref{sec:LCDM}, this shift is caused by the updated weak-lensing mass calibration using DES~Y3 data.
Finally, our constraint on $\gamma_\mathrm{ECS}$, the relative depth of the SPTpol~ECS survey, is centered on the value obtained from an SZ--richness scaling relation analysis (see Sec.~6.1.3 in \cite{bleem20}), but we recover a tighter constraint, most likely due to simultaneously analysing the cluster abundance and multi-observable scaling relations.

\begin{table}
  \caption{Constraints on the parameters of the observable--mass relations (mean and 68\% credible interval or 95\% limit).
  We do not show the prior-dominated parameters of the \Mwl--\Mhalo\ relation.
  Note that the constraints on the correlation coefficients are heavily affected by the prior $\mathcal U(-0.5,\,0.5)$.
  \label{tab:scaling_results}}
  \begin{ruledtabular}
    \begin{tabular}{lcc}
      Parameter & SPTclusters+lensing & SPTclusters+lensing\\
      && +{\it Planck}\\
      \colrule
      \multicolumn{2}{l}{SZ parameters} \\
      $\ln\asz$ & $0.72\pm0.09$ & $0.69\pm0.06$\\
      \bsz & $1.69\pm0.06$ & $1.73\pm0.04$\\
      \csz & $0.50\pm0.27$ & $0.74\pm0.11$\\
      \sigmalnzeta & $0.20\pm0.05$ & $0.21\pm0.05$\\
      $\gamma_\mathrm{ECS}$ & $1.05\pm 0.03$ & $1.05\pm0.03$\\
      \colrule
      \multicolumn{3}{l}{DES richness parameters (used for $z<1.1$)} \\
      $\ln\alambda$ & $3.74\pm0.07$ & $3.73\pm0.05$\\
      \blambda & $1.22\pm0.06$ & $1.25\pm0.04$\\
      \clambda & $-0.03\pm0.22$ & $0.15\pm0.12$\\
      \sigmalnlambda & $0.18\pm0.03$ & $0.18\pm0.04$\\
      \colrule
      \multicolumn{3}{l}{WISE richness parameters (used for $z>1.1$)} \\
      $\ln\alambda$ & $4.33\pm0.20$ & $4.33\pm0.21$\\
      \blambda & $0.93\pm0.09$ & $0.96\pm0.09$\\
      \clambda & $-2.1\pm0.6$ & $-2.0\pm0.6$\\
      \sigmalnlambda & $0.12\pm0.04$ & $0.12\pm0.05$\\
      \colrule
      \multicolumn{3}{l}{Correlation coefficients} \\
      $\rho_\mathrm{SZ,WL}$ & $<0.17$ & $<0.17$ \\
      $\rho_\mathrm{SZ,\tilde\lambda}$ & $<0.03$ & $<0.08$\\
      $\rho_\mathrm{WL,\tilde\lambda}$ & $-0.09\pm0.24$ & $-0.10\pm0.24$\\
    \end{tabular}
  \end{ruledtabular}
\end{table}

\subsubsection{The Richness--Mass Relation}

Throughout this work, we use richness measurements as computed by MCMF (see Sec.~\ref{sec:data}).
At low cluster redshift ($z_\mathrm{cluster}<1.1$), where our richness measurements are based on DES data, the recovered richness--mass relation implies that a halo of mass $M_{200\mathrm{c}}=3\times10^{14}~h^{-1}\Msun$ has a typical richness of 40.
We find a power-law dependence with mass that is steeper than unity at $>3\sigma$, and a dependence with redshift that is consistent with no evolution at $<1\sigma$.
The intrinsic scatter in $\ln\tilde\lambda$ is small ($0.18\pm0.03$), but note that the relation between observed richness $\lambda$ and mass contains an additional lognormal scatter of width $\tilde\lambda^{-1/2}$ [see Eq.~(\ref{eq:P_lambda_tildelambda})].\footnote{For intrinsic richness $\tilde\lambda<31$, the ``Poisson'' part of the uncertainty dominates over the intrinsic scatter.}

The parameters of the mean richness--mass relation at high redshift $z>1.1$, where we use WISE data, do not quite match those of the low-redshift richness--mass relation.
This is not entirely surprising, given how challenging it is to measure richness at these high redshifts and with the modest angular resolution of WISE.
Interestingly though, the intrinsic scatter is smaller than for DES-based richness at $1.5\sigma$.
The mass slope is consistent with unity within $1\sigma$, whereas the preferred redshift evolution is negative at the $\sim3\sigma$ level.
The significant difference between the richness measurements from the two observatories motivates the separate empirical calibration of each relation, as performed in this work.

We discuss our low-redshift richness--mass relation in the light of existing DES results.
These results are based on richness measurements using redMaPPer \citep{rykoff14}, whereas our richness measurements are based on MCMF, with $\mathrm{median}\left(\lambda_\mathrm{MCMF}/\lambda_\mathrm{redMaPPer}\right)=1.09$ and 24\% scatter \citep{klein19}.
The redMaPPer richness--mass relation was determined by cross-matching with SPT-confirmed clusters, and calibrating the SZ--mass relation assuming a fixed fiducial cosmology \citep{saro15, bleem20}, and through an analysis of the DES~Y1 cluster abundance with mass calibration enabled by cross-matching with SPT clusters, and using lensing data for SPT clusters \citep{costanzi21}.
Qualitatively, our richness--mass relation behaves as in these analyses, with residual differences being explained by the different richness measurements and by the fact that in this work, we calibrate the observable--mass relations using different lensing data (that is, DES~Y3).
The DES~Y1 weak-lensing based calibration of the redMaPPer richness--mass relation \citep{mcclintock19wl} was shown to yield a significantly shallower mass slope ($\blambda\approx1/1.356=0.74$) than the SPT-based analyses (e.g., $\blambda=1.020\pm0.080$ in \cite{bleem20}).
Within the limits of comparability, it appears that our constraint on the mass slope is in significant tension with the DES~Y1 lensing result \citep{mcclintock19wl}.

\subsubsection{The Correlation Coefficients of Intrinsic Scatter}
The constraints on the correlation coefficients (bottom rows in Table~\ref{tab:scaling_results}) are heavily affected by the uniform priors $\rho\sim\mathcal U(-0.5, 0.5)$ that we apply to make the covariance matrix nonsingular.
Note that as part of the blinding tests, we investigated the impact of fixing a set of fit parameters.
This included setting the correlation coefficients to 0.
The recovered scaling relation and cosmology constraints did not exhibit any significant shifts compared to the full analysis (see Fig.~\ref{fig:robustness}).
We therefore conclude that the exact range of the priors that we apply to the correlation coefficients does not have a significant impact on the recovered constraints on other parameters.

The posterior distributions of the two correlation coefficients that involve the SZ observable ($\rho_\mathrm{SZ,\tilde\lambda}$ and $\rho_\mathrm{SZ,WL}$) both peak at the lower bound $\rho=-0.5$ and we present the 95\% upper limits in Table~\ref{tab:scaling_results}.
A negative correlation coefficient $\rho_\mathrm{SZ,\tilde\lambda}<0$ is also found in the analysis of the LoCuSS cluster sample \citep{farahi19}, whereas simulation works tend to favor positive correlation coefficients between the SZ signal and other observables, e.g., \citep{stanek10, angulo12, shirasaki16}.
Negative correlation coefficients would imply that cluster triaxiality---which causes positive correlation---is not driving the joint scatter.
Our constraint on the correlation coefficient between lensing and richness ($\rho_\mathrm{WL,\tilde\lambda}$) peaks within the prior range, but the quoted constraints are also strongly affected by the prior.

In our hierarchical modeling of the cluster population, the correlation coefficients are degenerate with other parameters of the observable--mass relations.
Within our approach of empirical mass calibration, we do not expect strong constraints on the correlated scatter from current datasets (see also Sec.~4.1 in \cite{grandis21richnessSZ}).

\subsubsection{Observable--Mass Relations from Joint Analysis with {\it Planck} primary CMB}
In the last column in Table~\ref{tab:scaling_results}, we show the constraints obtained from the joint analysis of the cluster dataset and {\it Planck} primary CMB anisotropies.
Some parameters (\bsz, \csz, $C_{\lambda,\mathrm{DES}}$) shift within $1\sigma$ while most remain essentially unchanged.
This is explained by the fact that our cluster-based cosmology constraints overlap with the {\it Planck} results (see Fig.~\ref{fig:Om-sig8}).
Therefore, their combination does not pull the parameters away from their mean in the clusters-only analysis.
The uncertainties on most scaling relation parameters only tighten mildly when adding {\it Planck} CMB data.
However, there is a noticeable improvement on the constraints on the redshift evolutions \csz\ and $C_{\lambda,\mathrm{DES}}$.
These parameters are degenerate with \Om\ (see Fig.~\ref{fig:GTC_nuwCDM}), and the uncertainty on that parameter significantly tightens when adding {\it Planck} CMB data, as shown in Table~\ref{tab:results} and Fig.~\ref{fig:cl_planck_nuLCDM}.

\subsection{\wCDM\ with Massive Neutrinos}

\begin{figure}
  \includegraphics[width=\columnwidth]{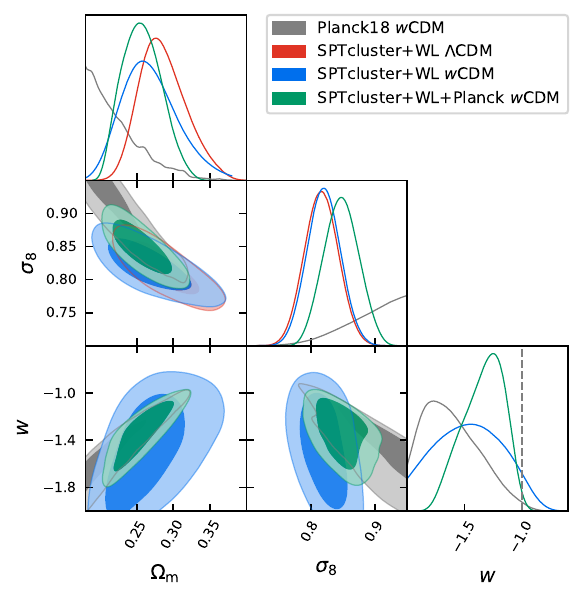}
  \caption{Constraints on \Om, \sig, and the dark energy equation of state parameter $w$ (68\% and 95\% credible regions) in a universe with massive neutrinos.
  We show results obtained from the analysis of the SPT cluster abundance, {\it Planck} primary CMB anisotropies, and their combination.
  The cluster dataset prefers a value that is more negative than a cosmological constant $w=-1$ at the $1.7\sigma$ level.
  Note that the constraints on \Om\ and \sig\ are only mildly affected by opening up $w$ as an additional degree of freedom.
  }
  \label{fig:wCDM}
\end{figure}

We consider an extension of the \LCDM\ model by opening up the dark energy equation of state parameter $w$ as an additional degree of freedom.
As before, we also marginalize over the sum of neutrino masses.
The constraints are summarized in Table~\ref{tab:results} and shown in Fig.~\ref{fig:wCDM}.
Considering only the cluster dataset, we note that the constraints on \Om\ and \sig\ are very similar to those obtained assuming the \LCDM\ model.
We obtain $w=-1.45\pm0.31$, a constraint that is more negative than that corresponding to a cosmological constant, with a probability to exceed of 0.10 ($1.7\sigma$) (note that the posterior distribution is affected by the hard cut $w>-2$).\footnote{Because the posterior distributions we recover for $w$ do not appear to be well-approximated by Gaussian distributions we compute the probability to exceed instead of applying Gaussian statistics.}
We point out that the uncertainty is comparable to the DES~Y3 3$\times$2~pt result ($w=-0.98^{+0.32}_{-0.20}$), underscoring the constraining power of cluster abundance measurements.
Figure~\ref{fig:wCDM} also shows wide degeneracies between \Om, \sig, and $w$ in the analysis of {\it Planck} primary CMB data.
In combination with the cluster dataset, some of these degeneracies can be broken, and the constraints are tightened.\footnote{Note that the sum of neutrino masses \sumMnu\ remains poorly constrained in $w$CDM even in the joint analysis.}
The joint constraints are qualitatively similar to those obtained from the cluster dataset alone, except for $w$, for which the uncertainty reduces significantly.
The mean recovered value for $w$ is less negative for the joint analysis, but the reduced uncertainty leads to a larger statistical difference with a cosmological constant (probability to exceed of 0.03, $2.2\sigma$) than for the clusters-only analysis.

Overall, the recovered constraints on $w$ are still quite weak.
As expected, $w$ exhibits a rather narrow degeneracy with the parameter of the SZ redshift evolution \csz\ (see Fig.~\ref{fig:GTC_nuwCDM}), which is only poorly constrained with the current weak-lensing data.
In \wCDM, we observe a shift to larger values ($\csz=1.00\pm0.32$), and away from its preferred value in \LCDM\ ($\csz=0.50\pm0.27$).
More powerful lensing data, especially at intermediate and high cluster redshifts, or a tight informative prior on the evolution of the ICM with redshift (as applied in the WtG analysis \citep{mantz15}) would enable tighter constraints on \csz, and thus on $w$.
Alternatively, a measurement of the gas fraction $f_\mathrm{gas}$ (as also applied in the WtG analysis) would constrain \Om\ independently, and through the \Om--$w$ parameter degeneracy, reduce the uncertainty on $w$.

\subsection{Modified Growth of Structure: $\sigma_8(z)$}

We consider a second, and final, extension of the \LCDM\ model, in which we allow the rate of structure growth to depart from the standard prediction.
We do so by directly measuring \sig\ at different redshifts.
Because the SPT cluster sample spans a wide range of redshifts, this exercise allows us to map the growth of cosmic structures since the universe was about 3~billion years old.

We define five bins in redshift such that each bin contains an equal number of clusters.
The six bin edges are $z\in\{0.25, 0.42, 0.57, 0.71, 0.93, 1.81\}$.
Then, we modify the computation of the matter power spectrum such that \sig\ at each bin center is normalized to one of five new fit parameters $\sigma_8(z_i)$ (see Table~\ref{tab:sigma8z}).
The observable--mass relations and the lensing data are treated the same as in the rest of this work, and we also marginalize over the sum of neutrino masses.
When analyzing the cluster dataset in this way, the recovered constraint on \Om\ degrades significantly compared to the \LCDM\ or \wCDM\ analyses.\footnote{The cluster abundance constrains a combination of \Om\ and \sig. Because the two parameters are degenerate, allowing more freedom for \sig\ weakens the constraint on \Om.}
Therefore, we consider an additional constraint on the matter density through the measurement of the sound horizon at recombination $100\,\theta_* = 1.04109\pm0.00030$ by {\it Planck} \citep{planck18VI}.\footnote{In combination with our priors on the Hubble parameter and the baryon density \Obhh, the measurement of $\theta_*$ enables a relatively loose constraint $\Om=0.295\pm0.065$.}

\begin{table}
  \caption{Constraints on \Om\ and $\sigma_8(z)$ for a \LCDM\ cosmology with massive neutrinos (mean and 68\% credible interval).
  Using the abundance of SPT clusters with DES~Y3 + HST weak-lensing mass calibration, and a prior on the sound horizon at recombination $\theta_*$ from {\it Planck}, we constrain \sig\ at discrete redshifts.
  We also quote $\sigma_8(z)$ for \LCDM\ and parameters from {\it Planck} 2018 TT,TE,EE+lowE.
  \label{tab:sigma8z}}
  \begin{ruledtabular}
    \begin{tabular}{lcc}
      Parameter & SPT cluster abundance+$\theta_*$ & {\it Planck}\\
      \colrule
$\Omega_\mathrm{m}$ & $0.320\pm0.029$ & $0.321\pm0.015$\\
$\sigma_8(z=0.34)$ & $0.670\pm0.026$ & $0.676\pm0.015$\\
$\sigma_8(z=0.49)$ & $0.599\pm0.022$ & $0.625\pm0.014$\\
$\sigma_8(z=0.64)$ & $0.603\pm0.023$ & $0.580\pm0.013$\\
$\sigma_8(z=0.82)$ & $0.524\pm0.020$ & $0.535\pm0.013$\\
$\sigma_8(z=1.35)$ & $0.406\pm0.017$ & $0.425\pm0.010$\\
    \end{tabular}
  \end{ruledtabular}
\end{table}

\begin{figure}
  \includegraphics[width=\columnwidth]{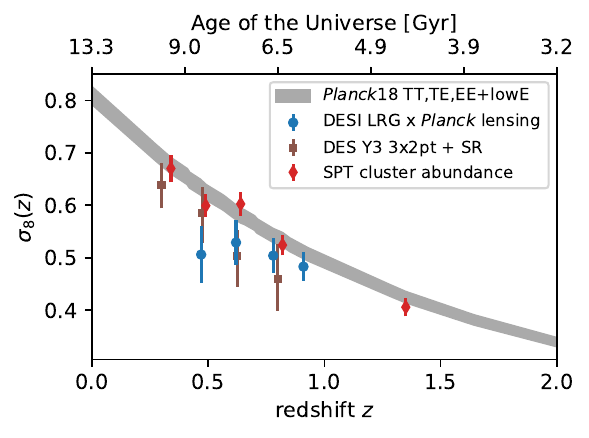}
  \caption{Evolution of $\sigma_8$ with time.
  Red diamonds show constraints from the SPT cluster abundance (mean and 68\% credible interval).
  Blue dots are obtained from the cross-correlation of DESI LRGs and {\it Planck} CMB lensing \citep{white22}.
  Both analyses include a {\it Planck}-based prior on the sound horizon at recombination $\theta_*$, with the value of $\theta_*$ fixed to the mean {\it Planck} value for DESI $\times$ {\it Planck}, and marginalized over the {\it Planck} posterior on $\theta_*$ for SPT clusters. 
  Brown squares show the constraints from DES Y3 3$\times$2~pt + shear ratios \citep{DESY3extensions}.
  The gray band shows the 68\% credible interval of the \LCDM\ prediction, assuming parameters determined by {\it Planck}.
  Our measurement of the growth of structure is consistent with that prediction.
  }
  \label{fig:sigma8_z}
\end{figure}

We summarize our results in Fig.~\ref{fig:sigma8_z}.
The gray band shows the 68\% credible interval of \sig\ as a function of redshift, assuming the \LCDM\ model with massive neutrinos, and parameters as measured by {\it Planck} primary CMB anisotropies.
Red data points show direct measurements of \sig\ at five redshifts, as probed by the cluster abundance in combination with the $\theta_*$ prior.
Note that the $\theta_*$ prior represents a subset of the information contained in the {\it Planck} CMB data, and we are thus comparing growth measurements for similar background cosmologies (thereby avoiding regimes where differences in $\sig(z)$ could be caused by differences in \Om\ instead of differences in structure growth).
Blue data points are obtained from a cross-correlation analysis of DESI LRGs and {\it Planck} CMB lensing, with $\theta_*$ fixed \citep{white22}.\footnote{Note that these measurements appear to be biased low due to degeneracies and parameter volume effects, as described in the original publication.}
Brown data points show the constraint from DES~Y3 galaxy clustering and lensing (3$\times$2~pt) and shear ratios \citep{DESY3extensions}.

Our measurements of $\sigma_8(z)$ presented in Table~\ref{tab:sigma8z} result in a difference with the \LCDM\ prediction with {\it Planck} parameters of $\chi^2=2.8$ for five parameters, which indicates good agreement.
However, the recovered parameter constraints are degenerate [the correlation coefficients between the individual $\sigma_8(z_i)$ are all $>0.76$], and taking these parameter correlations into account increases the $\chi^2$ value.
We anticipate that a more precise determination of the observable--mass relations, enabled by future lensing dataset, will reduce the level of correlation among the measured $\sigma_8(z_i)$ and will therefore enable more robust tests of the evolution of growth of structure.

\section{Summary}
\label{sec:summary}

In this work, we present cosmological constraints derived from the abundance of galaxy clusters selected in the SPT-SZ and SPTpol surveys \citep{bleem15, bleem20, klein23spt, bleem24}, and a simultaneous weak-lensing based mass calibration using DES~Y3 and HST data.
The cluster sample (1,005 confirmed clusters above $z>0.25$), the analysis of DES weak-lensing data for 688 sample clusters, the Bayesian analysis framework, and the validation of the analysis pipeline are presented in companion \citetalias{bocquet24I}.
The HST lensing measurements for 39 SPT clusters and their analysis were presented in earlier works \citep{schrabback18, raihan20, hernandez-martin20, schrabback21, zohren22}.

In a first phase of this work, we performed a blinded analysis in which we artificially offset the parameter constraints from their true output values by unknown \emph{blinding offset parameters} (see Sec.~\ref{sec:robustness_tests}).
This allowed us to perform a series of robustness tests with which we investigated the relative impact of alternative analysis setups on the results without knowing the actual parameter values.
These tests included various cuts in the cluster sample (excluding high-redshift and/or low-mass objects), details of the modeling of the SPT observable, and three alternative analyses of the DES~Y3 weak-lensing data.
Since none of these alternative analyses resulted in constraints that were significantly offset from the fiducial ones, we unblinded the results.
We did not further modify the data or the analysis, and we present the results as obtained after unblinding.

Assuming a flat \LCDM\ cosmology with massive neutrinos, we find that our constraints are consistent with, but tighter than, results obtained from previous weak-lensing calibrated cluster abundance measurements (see Fig.~\ref{fig:Om-sig8_cl} and Table~\ref{tab:results}).
Our constraints are competitive with other leading probes of the large-scale structure such as cosmic shear, joint 3$\times$2~pt analyses, and the CMB lensing power spectrum.
Our dataset best constrains the parameter combination $\sigma_8(\Om/0.3)^{0.25}=0.805\pm0.016$, with an uncertainty that is equal to the one recovered from {\it Planck} 2018 TT,TE,EE+lowE data.
The difference between our constraint on $S_8\equiv\sigma_8\sqrt{\Om/0.3}$ and the one from {\it Planck} is $1.1\sigma$, which does not strengthen the claim of an ``$S_8$ tension''.
The difference in the \Om--\sig\ plane has a probability to exceed of 0.47 ($0.7\sigma$).
The combination with the {\it Planck} dataset further tightens parameter constraints, and we place a 95\% upper limit on the sum of neutrino masses $\sumMnu<0.18$~eV.

Additionally allowing the dark energy equation of state parameter $w$ to vary does not change the recovered constraints on \Om\ and \sig\ in any significant way.
We recover $w=-1.45\pm0.31$.
The analysis of {\it Planck} primary CMB data exhibits a degeneracy between $w$ and \Om, \sig, and \sumMnu, which is broken in a joint analysis with our cluster dataset.
The resulting constraint on $w$ is still low, and we report $w=-1.34^{+0.22}_{-0.15}$, which differs from a cosmological constant by $2.2\sigma$.

Finally, we consider a phenomenological extension of the \LCDM\ model, in which we directly constrain the growth of structure.
Using the cluster dataset and a prior on the sound horizon at recombination $\theta_*$, we constrain \sig\ at five discrete redshifts between 0.34 and 1.35.
Our measurements of $\sigma_8(z)$ are consistent with the \LCDM\ prediction with {\it Planck} 2018 TT,TE,EE+lowE parameters (see Fig.~\ref{fig:sigma8_z} and Table \ref{tab:sigma8z}), but we note that our individual measurements are highly correlated.

A distinctive feature of this analysis is that, for the first time, we leverage the exquisite quality of weak-lensing data as provided by wide-field galaxy lensing surveys to perform a robust and precise weak-lensing informed measurement of the abundance of SZ-selected clusters.
In future work, we will use our cluster and lensing dataset to calibrate other cluster observables (X-rays, dynamics of cluster member galaxies) and to investigate further extensions of the \LCDM\ model, such as dark matter--dark radiation interactions and modified gravity models.
We also note that our analysis is the first cluster cosmology work that uses DES~Y3 lensing data.
This lensing dataset is also being used for mass calibration of redMaPPer-selected clusters in the DES~Y3 data and for mass calibration of further ICM-selected cluster samples.
Therefore, the DES lensing data play an important role in enabling cosmological constraints from the abundance of galaxy clusters.

Our measurements are still limited by size of the cluster sample, and by shape noise in the lensing measurements.
Therefore, increasing the cluster sample (thereby also increasing the number of lenses), increasing the size (and depth) of the lensing source sample, and the additional analysis of cluster lensing of the CMB e.g., \cite{baxter15, zubeldia&challinor19, raghunathan19DES}, will lead to improved cosmological constraints \citep{chaubal22}.
However, our analysis also highlights areas that deserve further attention (e.g., low-significance clusters, cluster member contamination, miscentering).
While the relative importance of these effects is currently at the $\lesssim1\sigma$ level (see Fig.~\ref{fig:robustness}), more work is needed to ensure that these do not become limiting factors in future analyses.
Opportunities for near-term improvements are numerous, with large ICM-selected cluster samples from the Atacama Cosmology Telescope \citep{hilton18}, eROSITA \citep{predehl21}, the SPT-3G survey \citep{benson14}, and the upcoming Simons Observatory \citep{ade19SO}.
These cluster samples will be analyzed jointly with a weak-lensing mass calibration using DES~Year~6 data, soon to be followed by galaxy and lensing data from Euclid \citep{laureijs11} and the Legacy Survey of Space and Time (LSST) \citep{lsst09sciencebook} at the Vera C. Rubin Observatory, whose combination is expected to be particularly powerful for extending sensitive cluster mass calibrations to higher redshifts \citep{rhodes17}.
Another promising avenue is to additionally consider the cluster 2-pt correlation function (and cluster--galaxy 2~pt), which further boosts the constraining power \citep{to21results, fumagalli24}.
Finally, the CMB-S4 survey will improve upon the aforementioned mm-wave surveys, and will extend analyses of SZ-selected cluster samples to well into the 2030s \citep{abazajian16cmbs4}.

\begin{acknowledgments}
This research was supported by the Excellence Cluster ORIGINS, which is funded by the Deutsche Forschungsgemeinschaft (DFG, German Research Foundation) under Germany's Excellence Strategy - EXC-2094-390783311, the MPG Faculty Fellowship program and the Ludwig-Maximilians-Universit\"at M\"unchen. Parts of the MCMC computations were carried out on the computing facilities of the Computational Center for Particle and Astrophysics (C2PAP). 
The Bonn and Innsbruck authors acknowledge support from the German Federal Ministry for Economic Affairs and Energy (BMWi) provided through DLR under projects 50OR2002 and 50OR2302, from the German Research Foundation (DFG) under grant 415537506, and the Austrian Research Promotion Agency (FFG) and the Federal Ministry of the Republic of Austria for Climate Action, Environment, Mobility, Innovation and Technology (BMK) via grants 899537 and 900565.
The Melbourne authors acknowledge support from the Australian Research Council’s Discovery Projects scheme (No. DP200101068).

The South Pole Telescope program is supported by the National Science Foundation (NSF) through the Grant No. OPP-1852617. Partial support is also provided by the Kavli Institute of Cosmological Physics at the University of Chicago.
PISCO observations were supported by US NSF grant AST-0126090. 
Work at Argonne National Laboratory was supported by the U.S. Department of Energy, Office of High Energy Physics, under Contract No. DE-AC02-06CH11357.

Funding for the DES Projects has been provided by the U.S. Department of Energy, the U.S. National Science Foundation, the Ministry of Science and Education of Spain, 
the Science and Technology Facilities Council of the United Kingdom, the Higher Education Funding Council for England, the National Center for Supercomputing 
Applications at the University of Illinois at Urbana-Champaign, the Kavli Institute of Cosmological Physics at the University of Chicago, 
the Center for Cosmology and Astro-Particle Physics at the Ohio State University,
the Mitchell Institute for Fundamental Physics and Astronomy at Texas A\&M University, Financiadora de Estudos e Projetos, 
Funda{\c c}{\~a}o Carlos Chagas Filho de Amparo {\`a} Pesquisa do Estado do Rio de Janeiro, Conselho Nacional de Desenvolvimento Cient{\'i}fico e Tecnol{\'o}gico and 
the Minist{\'e}rio da Ci{\^e}ncia, Tecnologia e Inova{\c c}{\~a}o, the Deutsche Forschungsgemeinschaft and the Collaborating Institutions in the Dark Energy Survey. 

The Collaborating Institutions are Argonne National Laboratory, the University of California at Santa Cruz, the University of Cambridge, Centro de Investigaciones Energ{\'e}ticas, 
Medioambientales y Tecnol{\'o}gicas-Madrid, the University of Chicago, University College London, the DES-Brazil Consortium, the University of Edinburgh, 
the Eidgen{\"o}ssische Technische Hochschule (ETH) Z{\"u}rich, 
Fermi National Accelerator Laboratory, the University of Illinois at Urbana-Champaign, the Institut de Ci{\`e}ncies de l'Espai (IEEC/CSIC), 
the Institut de F{\'i}sica d'Altes Energies, Lawrence Berkeley National Laboratory, the Ludwig-Maximilians-Universit{\"a}t M{\"u}nchen and the associated Excellence Cluster Origins, 
the University of Michigan, NSF's NOIRLab, the University of Nottingham, The Ohio State University, the University of Pennsylvania, the University of Portsmouth, 
SLAC National Accelerator Laboratory, Stanford University, the University of Sussex, Texas A\&M University, and the OzDES Membership Consortium.

Based in part on observations at Cerro Tololo Inter-American Observatory at NSF's NOIRLab (NOIRLab Prop. ID 2012B-0001; PI: J. Frieman), which is managed by the Association of Universities for Research in Astronomy (AURA) under a cooperative agreement with the National Science Foundation.

The DES data management system is supported by the National Science Foundation under Grant Numbers AST-1138766 and AST-1536171.
The DES participants from Spanish institutions are partially supported by MICINN under grants ESP2017-89838, PGC2018-094773, PGC2018-102021, SEV-2016-0588, SEV-2016-0597, and MDM-2015-0509, some of which include ERDF funds from the European Union. IFAE is partially funded by the CERCA program of the Generalitat de Catalunya.
Research leading to these results has received funding from the European Research Council under the European Union's Seventh Framework Program (FP7/2007-2013) including ERC grant agreements 240672, 291329, and 306478.
We  acknowledge support from the Brazilian Instituto Nacional de Ci\^encia e Tecnologia (INCT) do e-Universo (CNPq grant 465376/2014-2).

This work is based on observations made with the NASA/ESA {\it Hubble Space Telescope}, using imaging data from the SPT follow-up GO programs 12246 (PI: C.~Stubbs), 12477  (PI: F.~W.~High), 13412 (PI: T.~Schrabback), 14252 (PI: V.~Strazzullo), 14352 (PI: J.~Hlavacek-Larrondo), and 14677 (PI: T.~Schrabback).
STScI is operated by the Association of Universities for Research in Astronomy, Inc. under NASA contract NAS 5-26555.
It is also based on observations made with ESO Telescopes at the La Silla Paranal Observatory under programs 086.A-0741 (PI: Bazin), 088.A-0796 (PI: Bazin), 088.A-0889 (PI: Mohr), 089.A-0824 (PI: Mohr), 0100.A-0204 (PI: Schrabback), 0100.A-0217 (PI: Hern\'andez-Mart\'in), 0101.A-0694 (PI: Zohren),  and 0102.A-0189 (PI: Zohren).
It is also based on observations obtained at the Gemini Observatory, which is operated by the Association of Universities for Research in Astronomy, Inc., under a cooperative agreement with the NSF on behalf of the Gemini partnership: the National Science Foundation (United States), National Research Council (Canada), CONICYT (Chile), Ministerio de Ciencia, Tecnolog\'{i}a e Innovaci\'{o}n Productiva (Argentina), Minist\'{e}rio da Ci\^{e}ncia, Tecnologia e Inova\c{c}\~{a}o (Brazil), and Korea Astronomy and Space Science Institute (Republic of Korea), under programs  2014B-0338 and	2016B-0176 (PI: B.~Benson).

This manuscript has been authored by Fermi Research Alliance, LLC under Contract No. DE-AC02-07CH11359 with the U.S. Department of Energy, Office of Science, Office of High Energy Physics.
This research has made use of the SAO/NASA Astrophysics Data System and of adstex.\footnote{\url{https://github.com/yymao/adstex}}
We thank Ludwig, the SPT support cat for this analysis. Inquiries about SPT support cats shall be directed to T.C.
\end{acknowledgments}

\appendix
\section{Full Parameter Posterior Distributions}

\begin{figure*}
  \includegraphics[width=\textwidth]{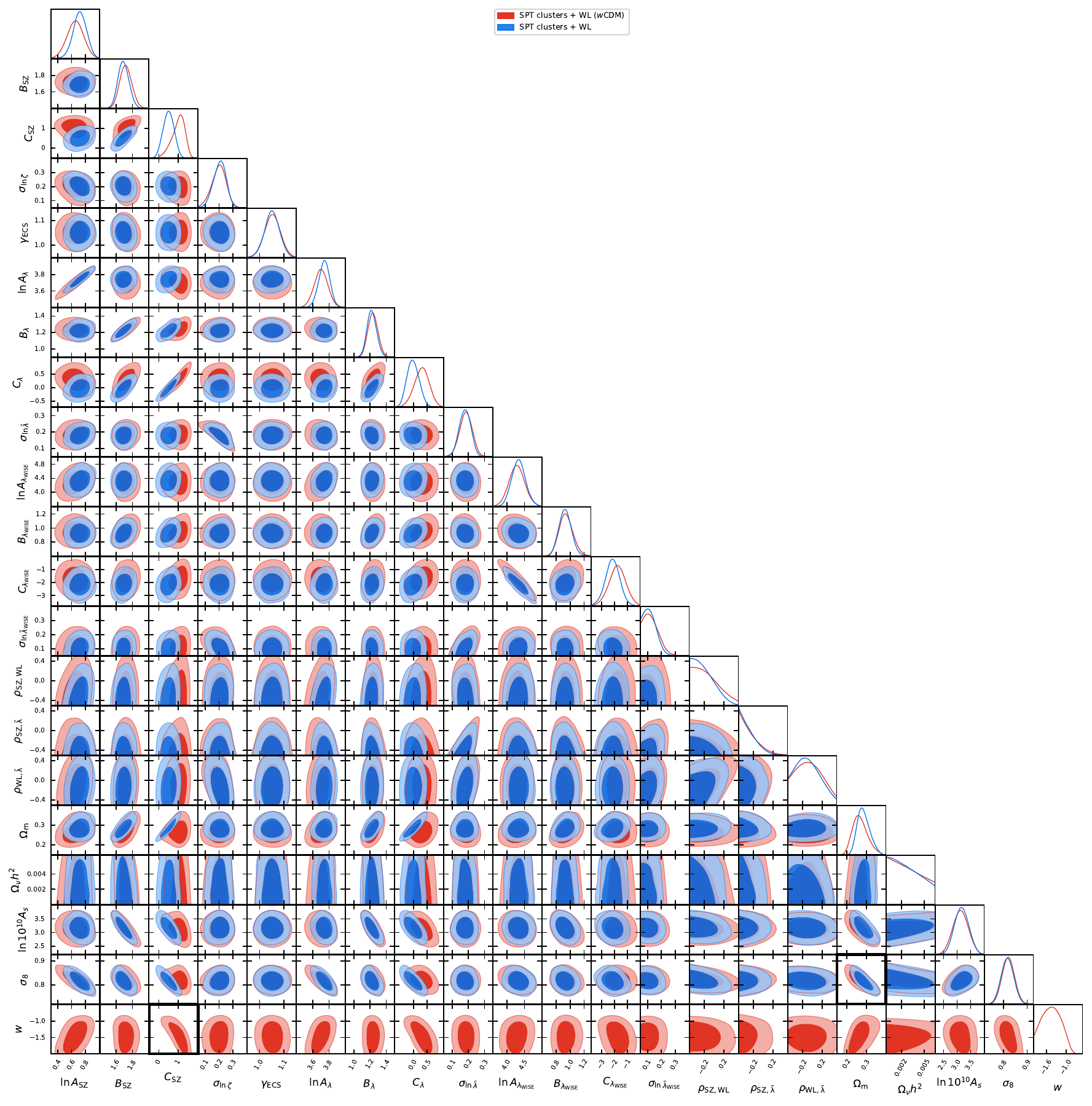}
  \caption{Parameter constraints on all parameters that are not strongly prior-dominated in the \LCDM\ and \wCDM\ analyses (68\% and 95\% credible regions).
  Between the two analyses, only the constraints on the redshift evolution parameters of the observable--mass relations \csz\ and \clambda\ shift by an appreciable amount.
  Conversely, the joint constraints on \Om\ and \sig\ are very similar (highlighted toward the bottom-right of the plot).
  The constraint on $w$ shows some significant degeneracies with other parameters, and in particular with the redshift evolution parameters \csz\ (highlighted at the bottom-left of the plot) of the SZ--mass relation, and \clambda\ of the richness--mass relation.
  }
  \label{fig:GTC_nuwCDM}
\end{figure*}

In Fig.~\ref{fig:GTC_nuwCDM}, we show the marginalized one- and two-dimensional parameter constraints obtained in the \LCDM\ and \wCDM\ analyses.
We show all parameters that are not strongly prior-dominated.
We highlight the joint constraint on \Om\ and \sig, which is quite robust to the extension with $w$ as a free parameter.
We also highlight the degeneracy between \csz\ and $w$, which implies that a tighter constraint on the redshift evolution of the observable--mass relation would enable tighter constraints on the dark energy equation of state.

\bibliographystyle{apsrev}
\bibliography{paperII}

\end{document}